%% file: ms.tex
\documentclass{article}

\pdfoutput=1

\usepackage[utf8]{inputenc}
\usepackage{newtxmath} 
\usepackage{newtxtext} 
\usepackage{fullpage}

\usepackage{authblk}

\usepackage[utf8]{inputenc}
\usepackage{newtxmath} 
\usepackage{newtxtext} 
\usepackage{amsmath}
\usepackage{fullpage}
\usepackage{subfig}
\usepackage{graphicx}
\usepackage[table]{xcolor}
\usepackage{tabularx}
\usepackage{bm}
\usepackage[disable]{todonotes}

\usepackage{color}
\usepackage{hyperref}

\usepackage{tikz}
\usetikzlibrary{external}
\newcommand{\dx}[0]{\Delta x}
\newcommand{\dy}[0]{\Delta y}

\newcommand{\partder}[2]{\frac{\partial #1}{\partial #2}}

\newcommand{\eastface}[1]{#1_{i + 1/2, j}}
\newcommand{\westface}[1]{#1_{i - 1/2, j}}

\newcommand{\cell}[1]{#1_{i,j}}

\newcommand{\eref}[1]{(\ref{#1})}

\makeatletter
\pretocmd\env@cases{\def\@rowc@lors{}}{}{}
\pretocmd\start@align{\def\@rowc@lors{}}{}{}
\pretocmd\start@array{\def\@rowc@lors{}}{}{}
\pretocmd\start@split{\def\@rowc@lors{}}{}{}
\makeatother    

\rowcolors{2}{gray!15}{}

\newcommand{\secref}[1]{Section~\ref{#1}}

\newcommand{\havahol}[1]{\todo[color=yellow!40]{H{\aa}vard: #1}}

\input{tikz/tikzsetup.tex}

\begin{document}

\title{Real-World Oceanographic Simulations on the GPU\\using a Two-Dimensional Finite-Volume Scheme}


\author[1,2]{Andr\'{e}~R.~Brodtkorb\footnote{Corresponding author: Andre.Brodtkorb@met.no}}
\author[3,4]{H\aa{}vard~Heitlo~Holm}

\affil[1]{
Norwegian Meteorological Institute, 
P.O.~Box 43~Blindern,
NO-0313~Oslo,
Norway.}
\affil[2]{
Oslo Metropolitan University,
Department of Computer Science,
P.O.~Box 4~St. Olavs plass,
NO-0130~Oslo,
Norway.
}
\affil[3]{
SINTEF~Digital,
Mathematics~and~Cybernetics,
P.O.~Box 124~Blindern,
NO-0314~Oslo,
Norway.}
\affil[4]{
Norwegian~University~of~Science~and~Technology,
Department~of~Mathematical~Sciences,
NO-7491~Trondheim,
Norway.}

\date{}

\maketitle

\begin{abstract}
In this work, we take a modern high-resolution finite-volume scheme for solving the rotational shallow-water equations and extend it with features required to run real-world ocean simulations. 
Our contributions include a spatially varying north vector and Coriolis term required for large scale domains, moving wet-dry fronts, a static land mask, bottom shear stress, wind forcing, boundary conditions for nesting in a global model, and an efficient model reformulation that makes it well-suited for massively parallel implementations.
Our model order is verified using a grid convergence test, and we show numerical experiments using three different sections along the coast of Norway based on data originating from operational forecasts run at the Norwegian Meteorological Institute. 
Our simulation framework shows perfect weak scaling on a modern P100 GPU, and is capable of providing tidal wave forecasts that are very close to the operational model at a fraction of the cost.
All source code and data used in this work are publicly available under open licenses. 
\end{abstract}

\section{Introduction}
The aim of this paper is to simulate realistic oceanographic scenarios using a modern finite-volume scheme on GPUs. 
Modern operational ocean models, such as the Regional Ocean Modeling System (ROMS)~\cite{roms} and the Nucleus for European Modelling of the Ocean (NEMO)~\cite{nemo}, are based on solving the primitive equations that describe conservation of mass, momentum, temperature, and salinity.
Even though these ocean models are heavily optimized, their complex physical and mathematical properties make them very computationally demanding.
The NorKyst-800 model system~\cite{norkyst800_2011}, for example, is based on the ocean model ROMS and covers the Norwegian coast with 800~m horizontal resolution.
It is run operationally by the Norwegian Meteorological Institute on a daily basis.
The complete model consists of 42 vertical layers with $2600\times900$ grid cells each, and a 24 hour forecast takes 45 minutes when using 256 CPUs~\cite{kaihc}.
It is unfeasible to use such models for flexible on-demand simulations during search-and-rescue operations, oil spills, heavy lifting at sea, etc. In these cases, fast on-demand simulations based on the shallow-water equations can have a significant value for decision makers.

We simulate the shallow-water equations in a rotating frame of reference, targeting efficient simulation of ocean currents based on the operational NorKyst-800 data.
Our model equations are essentially two-dimensional simplifications of the primitive equations and assume constant density and a vertically integrated momentum (see more detailed derivations in, e.g., R{\o}ed~\cite{lproed_2019}).
Our work starts with the recent high-resolution finite-volume scheme by Chertock et al.~\cite{Chertock2017}, and extends this scheme to support bottom shear stress, wind forcing, projection-free spatial variations in the north vector, spatially varying Coriolis parameter, land mask, and a moving wet-dry boundary.
We present how to reconstruct a piecewise bilinear bathymetry from the piecewise constant bathymetry in the NorKyst-800 data, nesting of the model into the NorKyst-800 data using linearly interpolated boundary conditions, and generation of both higher and lower spatial resolution of the initial conditions.

We simulate increasingly challenging scenarios along the Norwegian coast, and compare with reference results from the operational model. As the shallow-water equations are particularly well suited for tidal and storm-surge predictions, we also compare our tidal-wave predictions at several locations with the reference dataset and observations.
All datasets and source code used in this work are publicly available under open licenses.%
\footnote{%
The source code used to create plots and results in this article is available in revision \texttt{7281301a3a286351b6e857add12c2fbd0165b229} of the \texttt{gpu-ocean} repository, \url{https://github.com/metno/gpu-ocean}, released under GPL version 3. The plots are created using Jupyter Notebooks~\cite{Kluyver:2016aa} in the folder \texttt{demos/realisticSimulations}, and the simulator code itself is under the directory \texttt{SWESimulators} as a Python module.
We have used the NorKyst-800 forecast for 2019-07-16 from the Norwegian Meteorological Institute in this work, available on\\ \url{https://thredds.met.no/thredds/dodsC/fou-hi/norkyst800m-1h/NorKyst-800m_ZDEPTHS_his.an.2019071600.nc}. The observed sea-surface levels are obtained from Se Havniv{\aa}, \url{https://www.kartverket.no/sehavniva/}, provided by the Norwegian Mapping Authority, Hydrographic Service.
The data sets are available under the Creative Commons 4.0 BY International license as described on \url{https://www.met.no/en/free-meteorological-data/Licensing-and-crediting} and 
\url{https://www.kartverket.no/en/data/Terms-of-use/}.%
}
\havahol{Cite the future zenodo release of the code}

The rest of the paper is organized as follows:
Section~\ref{sec:mathematics} gives a high-level description of the shallow-water equations and how they can be solved numerically.
In Section~\ref{sec:sim_extensions}, we present all the extensions that are required to tackle relevant oceanographic problems.
The performance and numerical accuracy of our scheme are assessed in Section~\ref{sec:performance_evaluation}, where we show perfect weak scaling of the GPU implementation and run a grid convergence test with the new terms introduced in the scheme.
We present operational-grade simulations in Section~\ref{sec:real_world_sims}, and finally summarize the work in Section~\ref{sec:summary}.

\section{Mathematical formulation}
\label{sec:mathematics}
We use the shallow-water equations with source terms to simulate ocean dynamics.
Since the equations are purely barotropic, they are not expected to generate or preserve features such as mesoscale eddies in the long term. 
These features arise from baroclinic instabilities, typically caused by variations in temperature and salinity that are not included in the mathematical model, but
the equations are nevertheless capable of capturing the most important physics required to simulate short-term ocean dynamics.
We can write the equations with source terms as
\begin{equation}
\rowcolors{2}{}{} 
    \begin{split}
    \left[
    \begin{array}{c}
    h\\
    hu\\
    hv
    \end{array}
    \right]_t
    +
    \left[
    \begin{array}{c}
    hu\\
    hu^2+\tfrac{1}{2}gh^2\\
    huv
    \end{array}
    \right]_x
    &+
    \left[
    \begin{array}{c}
    hv\\
    huv\\
    hv^2+\tfrac{1}{2}gh^2
    \end{array}
    \right]_y
    \\
    \,=&
    \left[
      \begin{array}{c}
        0 \\
        ghH_x \\
        ghH_y
      \end{array}
    \right]
  +
    \left[
      \begin{array}{c}
        0 \\
        -ru\sqrt{u^2+v^2}/h \\
        -rv\sqrt{u^2+v^2}/h
      \end{array}
    \right]
  +
    \left[ 
      \begin{array}{c}
      0 \\ 
      fhv \\ 
      -fhu 
      \end{array} 
    \right]
  +
    \left[ 
      \begin{array}{c}
      0 \\ 
      \tau_x \\ 
      \tau_y 
      \end{array} 
    \right].
    \end{split}
    \label{eq:swe}
\end{equation}
Here, $h$ is water depth, $hu$ momentum along the $x$-axis, and $hv$ momentum along the $y$-axis (see also Figure~\ref{fig:sweoverview}). 
Furthermore, $g$ is the gravitational acceleration, $H$ is the equilibrium ocean depth measured from a reference sea-surface level, $r$ is the friction coefficient,
$f$ is the Coriolis parameter, and $\tau_x$ and $\tau_y$ are the wind stress along the $x$- and $y$-axis, respectively. In vector form, we can write the equations as
\begin{equation}
    Q_t + F(Q)_x + G(Q)_y = B(Q) + S(Q) + C(Q) + W(Q).
    \label{eq:swe_vec}
\end{equation}
Here, $Q = [h, hu, hv]^T$ is the vector of conserved variables and $B$, $S$, $C$, and $W$ are source terms representing varying bathymetry, bed shear stress, Coriolis force, and wind drag, respectively. 

By using a finite-volume discretization on a Cartesian grid to solve \eqref{eq:swe_vec}, we get the following semi-discrete formulation
\begin{align}
\frac{\partial Q}{\partial t} &= 
  - \frac{1}{\Delta x}\bigl[F_{i+1/2,j} - F_{i-1/2,j}\bigr]
  - \frac{1}{\Delta y}\bigl[G_{i,j+1/2} - G_{i,j-1/2}\bigr]
  + B_{i, j} + S_{i, j} + C_{i, j} + W_{i, j}.
  \label{eq:sw_semidiscrete}
\end{align} 
To evolve the solution in time, we use a second-order strong stability-preserving Runge-Kutta scheme~\cite{SSP_gottlieb_shu_tadmor_2001},
\begin{equation}
    \begin{split}
    Q^*_{i, j} &= Q^n_{i, j} + \Delta t M(Q^n)_{i, j}\\
    Q^{n+1}_{i, j} &= \tfrac{1}{2} \Bigl(Q^n_{i, j} + \left(Q^*_{i, j} + \Delta t M(Q^*)_{i, j}\right) \Bigr),
    \end{split}
    \label{eq:rk2}
\end{equation}
in which $M$ represents the right hand side of \eref{eq:sw_semidiscrete}.
The time step is restricted by the CFL condition given by
\begin{equation}
  \Delta t 
  \leq 
  \frac{1}{4} 
  \text{min} \left\{
  \frac{\Delta x}{\text{max} \left| u\pm\sqrt{gh} \right|}, 
  \frac{\Delta y}{\text{max}\left| v\pm\sqrt{gh} \right|}
  \right\},
  \label{eq:cfl}
\end{equation}
if we assume no Coriolis and wind drag. 
The dominant term for oceanographic applications is typically the speed of gravity waves $\sqrt{gh}$, and we use a conservative Courant number less than one in our simulations to account for Coriolis and wind drag.

\subsection{Numerical scheme}
The starting point for our simulator is the scheme proposed by Chertock et al.~\cite{Chertock2017}, from now on referred to as CDKLM. It is a modern central-upwind scheme that uses a reconstruction of the physical variables to be able to capture steady-state solutions important for physical oceanography, and includes source terms for varying bathymetry and Coriolis. The original paper includes several idealized test cases and examples demonstrating that the steady states in geostrophic balance are well captured. However, there are several shortcomings that need to be addressed before the scheme can be used for real-world simulations.

\begin{figure}
    \begin{center}
        \begin{centering}
        { 
          \subfloat[ ]{\input{tikz/equation.tikz}}
          \hspace{1em}
          \subfloat[ ]{\input{tikz/equation_discrete.tikz}}
        }
        \end{centering}
        \\
        \begin{centering}
        {
          \subfloat[ ]{\input{tikz/equation_slopes.tikz}}
          \hspace{1em}
          \subfloat[ ]{\input{tikz/equation_point_values.tikz}}
        }
        \end{centering}
    \caption{a) The relationship between the physical variables $h$, $H$, $\eta$, and $hu$ used in the shallow-water equations, here shown in one dimension. b) The ocean state is discretized in terms of cell average values, whereas the bathymetry is defined at the cell intersections. c) We find the slopes of the ocean state within each cell.
    d) The flux between the cells is found from two one-sided point values at each interface, as seen from the two adjacent cells.}
    \label{fig:sweoverview}
    \end{center}
\end{figure}
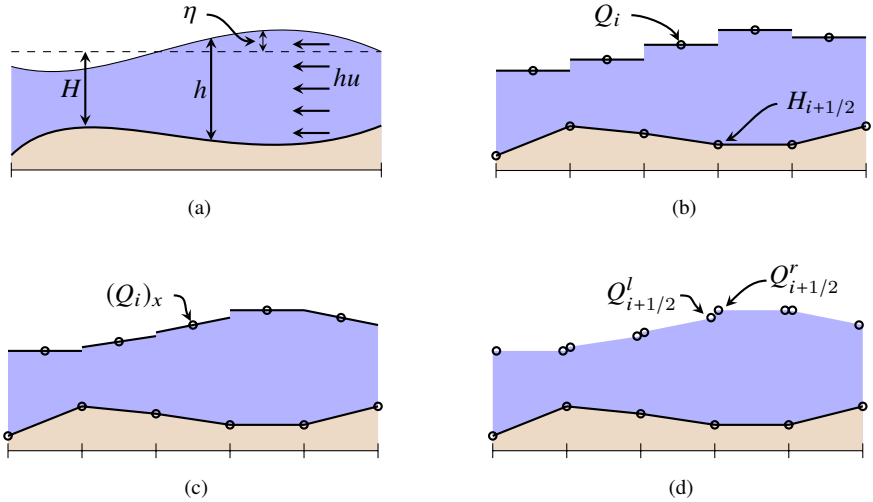

The CDKLM scheme is based on the same principles as the MUSCL scheme~\cite{vanLeer79} to obtain a high-order, total variation diminishing discretization. For each cell in the domain, we first reconstruct a piecewise (linear) function for our physical variables (see Figure~\ref{fig:sweoverview}). 
At the interface between two cells, we evaluate the slope limited function from the right and left cell, and use these two values to compute the flux across the interface. The scheme uses the Riemann-solver-free central-upwind flux function,
\begin{align}
F_{i+1/2} = \frac{a^+ F(Q_{i+1/2}^l) - a^- F(Q_{i+1/2}^r)}{a^+ - a^-}
+ \frac{a^+ a^-}{a^+ - a^-}\left[Q_{i+1/2}^r - Q_{i+1/2}^l \right],
\label{eqn:numerical_flux}
\end{align}
to evaluate the numerical flux~\cite{kurganov_noelle_petrova_2001}.
Here, $a^+$ and $a^-$ are the largest positive and negative wave speeds at the interface, respectively. 

When adding source terms, it becomes challenging to preserve steady states such as ``lake at rest'', i.e., 
\begin{equation}
    \partder{Q}{t} = 0 \qquad \mathrm{if} \qquad 
    \partder{\eta}{x} = \partder{\eta}{y} = 
    hu = hv = 0,
    \label{eq:lakeatrest}
\end{equation}
in which $\eta$ is the sea-surface deviation from a mean equilibrium level, given by
\begin{equation}
    \eta = h - H.
    \label{eq:eta}
\end{equation}
Kurganov and Levy~\cite{kurganovLevy2002} presented a discretization of the bottom slope source term $B(Q)$ and a matching reconstruction based on water elevation to capture such steady states. For oceanographic simulation, an important steady state is the so-called geostrophic balance, in which the angular momentum caused by the rotational reference frame is balanced by the pressure gradient:
\begin{equation}
    \begin{split}
        f hv - gh \frac{\partial \eta}{\partial x} = 0, \qquad
        -f hu - gh \frac{\partial \eta}{\partial y} = 0.
    \end{split}
    \label{eq:geostrophicBalance}
\end{equation}
The novel idea in the CDKLM scheme is to reconstruct the face values $Q^l_{i+1/2}$ and $Q^r_{i+1/2}$ based on this geostrophic balance so that the resulting flux $F_{i+1/2}$ also balances the Coriolis force for steady-state solutions. 
This makes the scheme suitable for simulating ocean currents in a rotating frame of reference.

\section{Efficient simulation of real-world ocean currents}
\label{sec:sim_extensions}
Our simulator is implemented on the GPU for efficiency. Explicit finite-volume schemes normally have an embarrassingly parallel nature, and it is therefore natural to implement them on such massively parallel accelerator hardware (see e.g.,~\cite{AMC10:JS, sw12, 2018_parna_weno_swe_gpu, 2019_qin_gpu_geoclaw}). 
We use CUDA~\cite{cuda_programming_guide} coupled with PyCUDA~\cite{kloeckner_pycuda_2012} to access the GPU from Python in this work, and 
we have previously shown that using Python instead of C++ has a negligible performance impact when used like this~\cite{holm_pycuda_mdpi}. 
Using the Python ecosystem for tasks such as pre-processing input data and post-processing results has significantly increased our productivity, while still maintaining high performance.

A simulation starts by reading initial conditions, forcing terms, and other grid data from NetCDF files hosted by \emph{thredds} servers at the Norwegian Meteorological Institute. We continue by initializing the simulator, which internally allocates memory and uploads initial conditions on the GPU. Finally, the main loop steps the simulator forward in time, and occasionally (e.g., every simulated hour) downloads and writes results to NetCDF files on disk. 

The rest of this section covers additions and improvements to the CDKLM scheme required for simulating real-world scenarios.

\subsection{Efficient parallel formulation of CDKLM}
\label{sec:reformulation}
The original CDKLM scheme uses a recursive formulation of the potential energies used for the well-balanced flux reconstruction.
This turns the reconstruction procedure into a global operation with a data-dependency spanning the whole domain, making the scheme prohibitively expensive on parallel architectures.
The recursive expression can be calculated somewhat efficiently using a prefix sum, as suggested by the authors themselves,
but by carefully reformulating the recursive terms it can also be rephrased as a local operation. 

The scheme reconstructs $h$, $hu$, and $hv$ on each side of a face to calculate the fluxes in \eqref{eqn:numerical_flux}.  
However, to capture the rotating steady-state solutions in geostrophic balance, the reconstruction is based on the potential energies,  
\begin{equation}
	K = g(\eta - V), \qquad L = g(\eta + U),
	\label{eq:potentialEnergies}
\end{equation}
instead of physical variables. 
Here, $U$ and $V$ are the primitives of the Coriolis force, given by
\begin{equation} 
	V_x = \frac{f}{g}v, \qquad U_y = \frac{f}{g}u.
	\label{eq:coriolisPrimitives}
\end{equation}
Note that setting $K_x = 0$ and $L_y = 0$ is equivalent to the geostrophic balance from \eqref{eq:geostrophicBalance}.
In the following, we detail the reconstruction along the abscissa, but the same derivations apply in the ordinate.
The recursive terms appear from the discretization of $U$ and $V$, which are defined on the cell faces as
\begin{equation}
	\eastface{V} = \westface{V} + \frac{\dx}{g}f_{i,j}v_{i,j}, \qquad V_{-1/2, j} = 0.
	\label{eq:recursiveV}
\end{equation}
The Coriolis parameter $f_{k,j}$ is allowed to change both along the $x$ and $y$ axis, even though $f$ only varies with latitude.
This is because we will use a spatially varying latitude, as described in more detail in Section~\ref{sec:coriolis}.
Values of $V$ in the cell centers are obtained by taking the mean of the face values, so that the cell-average value of $K$ from  \eqref{eq:potentialEnergies} becomes
\begin{equation}
	K_{i,j} = g \left(\eta_{i,j} - \frac{1}{2} \left(\eastface{V} + \westface{V} \right)\right).
	\label{eq:discK}
\end{equation}
By reconstructing the (limited) slopes of $K$ within each cell, the face values of $h$ are found by combining \eref{eq:discK} and \eref{eq:eta}, so that
\begin{equation}
		h_{i,j}^E = \frac{1}{g} \left( K_{i,j} + \frac{\dx}{2}(K_x)_{i,j} \right) + \eastface{V} + \eastface{H} , 
	\label{eq:hOnFaces_E}
\end{equation}
and
\begin{equation}
		h_{i,j}^W = \frac{1}{g} \left( K_{i,j} - \frac{\dx}{2}(K_x)_{i,j} \right) + \westface{V} + \westface{H}.
	\label{eq:hOnFaces_W}
\end{equation}

The reconstructed values \eref{eq:hOnFaces_E} and \eref{eq:hOnFaces_W} of $h$ at the cell faces quickly become a performance bottleneck as long as they depend on calculating the recursive relation in \eqref{eq:recursiveV}.
As it turns out, however, we never need to explicitly compute the recursive terms.
First of all, notice that \eref{eq:hOnFaces_E} and \eref{eq:hOnFaces_W} only depend on derivatives of $K$ and not on $L$.
Similarly, the reconstructions of $\cell{h}^N$ and $\cell{h}^S$ depend on $L$ and $L_y$.
This means that we are only interested in derivatives of $K$ (and $L$) in the same direction as the recursive terms for $V$ (and $U$).
The derivatives are limited using the generalized minmod function using the backward, central, and forward differences.
The backward difference is given by
\begin{equation}
	\frac{K_{i,j} - K_{i-1,j}}{\dx}  = 
	    \frac{g}{\dx} \left(
            \eta_{i,j} - \frac{1}{2} \left(\eastface{V} + \westface{V} \right) 
            - \eta_{i-1,j} + \frac{1}{2} \left( \westface{V} + V_{i- 3/2, j} \right)
        \right).
	\label{eq:Kbackward1}
\end{equation}
We see that $\westface{V}$ cancels, and focus on the remaining $V$ values. By applying the recursive expression in \eqref{eq:recursiveV} twice on $\eastface{V}$ we get
\begin{equation}
	\begin{split}
		\frac{1}{2}\left(- V_{i+1/2, j} + V_{i- 3/2, j} \right)
		=& \frac{1}{2} \left( - V_{i-1/2, j} - \frac{\dx}{g}f_{i,j}v_{i,j} + V_{i- 3/2, j} \right) \\
		=& \frac{1}{2} \left( - V_{i-3/2, j} - \frac{\dx}{g}f_{i-1,j}v_{i-1, j} - \frac{\dx}{g}f_{i,j}v_{i,j} + V_{i- 3/2, j} \right) \\
		=\,&  - \frac{\dx}{2g} \Big( f_{i-1,j}v_{i-1,j} + f_{i,j} v_{i,j} \Big).
	\end{split}
	\label{eq:Kbackward2}
\end{equation}
Inserted back into \eqref{eq:Kbackward1}, the backward difference needed to evaluate $(K_{i,j})_x$ can be written as
\begin{equation}
	\frac{K_{i,j} - K_{i-1,j}}{\dx} = \frac{g}{\dx}\left(\eta_{i,j} - \eta_{i-1,j} - \frac{\dx}{2g} \Big( f_{i-1,j}v_{i-1,j} + f_{i,j} v_{i,j} \Big) \right),
	\label{eq:Kbackwards}
\end{equation}
which no longer contains any recursive terms.
Similar derivations can be used for the forward and central differences, and equivalently to obtain $(L_{i,j})_y$.

After obtaining  $(K_x)_{i,j}$ we look at the reconstruction of $h_{i,j}^E$ from \eqref{eq:hOnFaces_E}.
We start by inserting the expression for $K$ from \eref{eq:discK} into \eqref{eq:hOnFaces_E} and gather all $V$-terms,
\begin{equation}
	\begin{split}
	h_{i,j}^E 
			&= \eta_{i,j} - \frac{1}{2} \left(\eastface{V} + \westface{V} \right) + \frac{\dx}{2g}(K_x)_{i,j} + V_{i+1/2,j} + H_{i+1/2,j} \\
			&= \eta_{i,j} + H_{i+1/2,j} + \frac{\dx}{2g}(K_x)_{i,j} + \frac{1}{2}\left( V_{i+1/2,j} - V_{i-1/2,j} \right).
	\label{eq:hE1}
	\end{split}
\end{equation}
Here, the values for $\eta_{i,j}$, $H_{i,j}$ and $(K_x)_{i,j}$ are known, and by using the recursive expression in \eqref{eq:recursiveV} we get that
\begin{equation}
	\begin{split}
        \frac{1}{2}\left( V_{i+1/2,j} - V_{i-1/2,j} \right) &= \frac{1}{2}\left(  V_{i-1/2,j}  + \frac{\dx}{g}v_{i,j}f_{i,j} - V_{i-1/2,j} \right) \\
		&= \frac{\dx}{2g}f_{i,j}v_{i,j}.
	\end{split}
	\label{eq:hE2}
\end{equation}
Inserting \eqref{eq:hE2} back into \eref{eq:hE1} we see that we get
\begin{equation}
	h_{i,j}^E = \eta_{i,j} + H_{i+1/2,j} + \frac{\dx}{2g}\Big[ (K_x)_{i,j} + f_{i,j} v_{i,j} \Big],
	\label{eq:hE}
\end{equation}
Again, the same derivation can be applied on the other three face values as well.
This enables us to express the reconstruction step in terms of only local variables, which unlocks an embarrassingly parallel numerical algorithm.

\subsection{Accuracy, precision and GPUs}
\label{sec:accuracy}
Today's GPUs can be categorized in two major classes: GPUs meant for the consumer market, and GPUs meant for supercomputers and other professional users. For the most part, these GPUs are close to identical, but there are some key differences. One is price: professional GPUs can cost over ten times more than equivalent consumer market GPUs, making the latter tractable from a financial stand point. Another difference is in the feature set. Professional GPUs offer more memory, error-correcting code (ECC) memory, and significantly higher performance for features such as double precision. 

Double precision is not an important feature on consumer market GPUs, and typically offer just 3\% of the performance of single-precision arithmetics. The NVIDIA TITAN RTX, for example, has a performance of 16.3 teraFLOPS in single precision, and 0.51 teraFLOPS in double precision. On professional GPUs, on the other hand, the performance is 50\% of single precision, which is the same ratio as on CPUs. If we are able to utilize single-precision calculations, we should therefore be able to get the highest performance at minimum monetary cost. 
We have previously shown that using single precision is sufficiently accurate for most simulation scenarios for the shallow-water equations~\cite{brodtkorb2010simulation}, and recent studies have shown a 40\% increase in performance for complex forecasting models when using single precision~\cite{vavna2017single}.

It is well known that floating-point numbers have a round-off error that increases for larger numbers. A single-precision floating-point number is represented as
\begin{align}
    (-1)^s \cdot 2^{e-127} \cdot (1.0 + f) 
\end{align}
in which $s$ is the sign bit, $e$ is represented using 8 bits, and $f$ is the fractional part, represented by 23 bits. This formulation means that the distance between two floating-point numbers is smallest close to zero, and increases as $e$ increases. This is important to note in oceanographic simulations, as the water depth, $h$, can often be thousands of meters, whereas the relevant tidal wave height is perhaps 0.5 meters. The loss in precision due to the water depth makes floating point prohibitively inaccurate if we implement the shallow-water equations as formulated in \eref{eq:swe}. To counter this loss in precision, we can possibly use double precision (at twice the computational and memory cost), but a better alternative is to reformulate the problem to represent the relevant physical quantities. 
This can be done by basing our simulation on $\eta$ from \eqref{eq:eta} instead of $h$ and use $[\eta, hu, hv]^T$ as our vector of conserved variables.
By combining this approach with single-precision floating-point numbers, we get more than sufficient accuracy for the relevant oceanographic simulations. 

\begin{figure}
    \centering
    \subfloat{\includegraphics[width=0.45\textwidth]{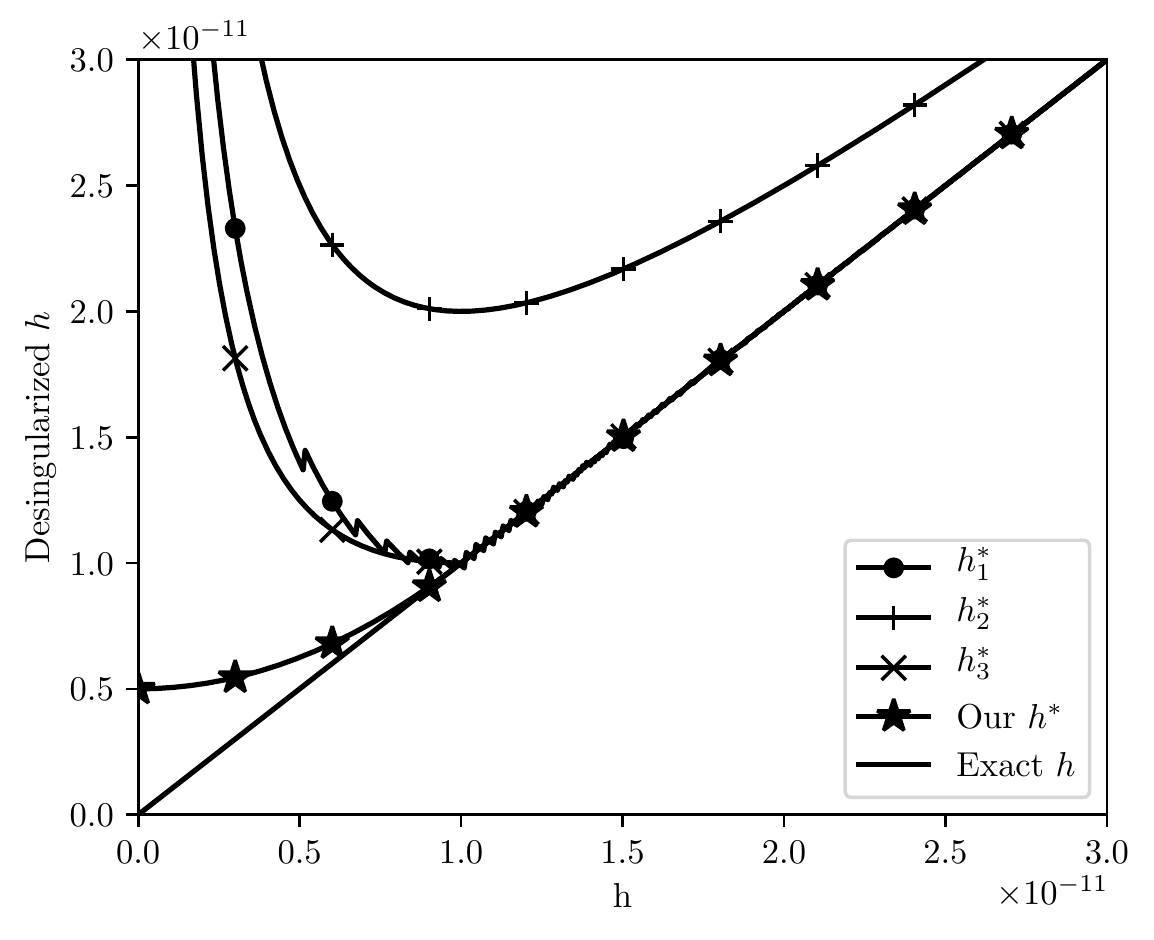}}\qquad
    \subfloat{\includegraphics[width=0.45\textwidth]{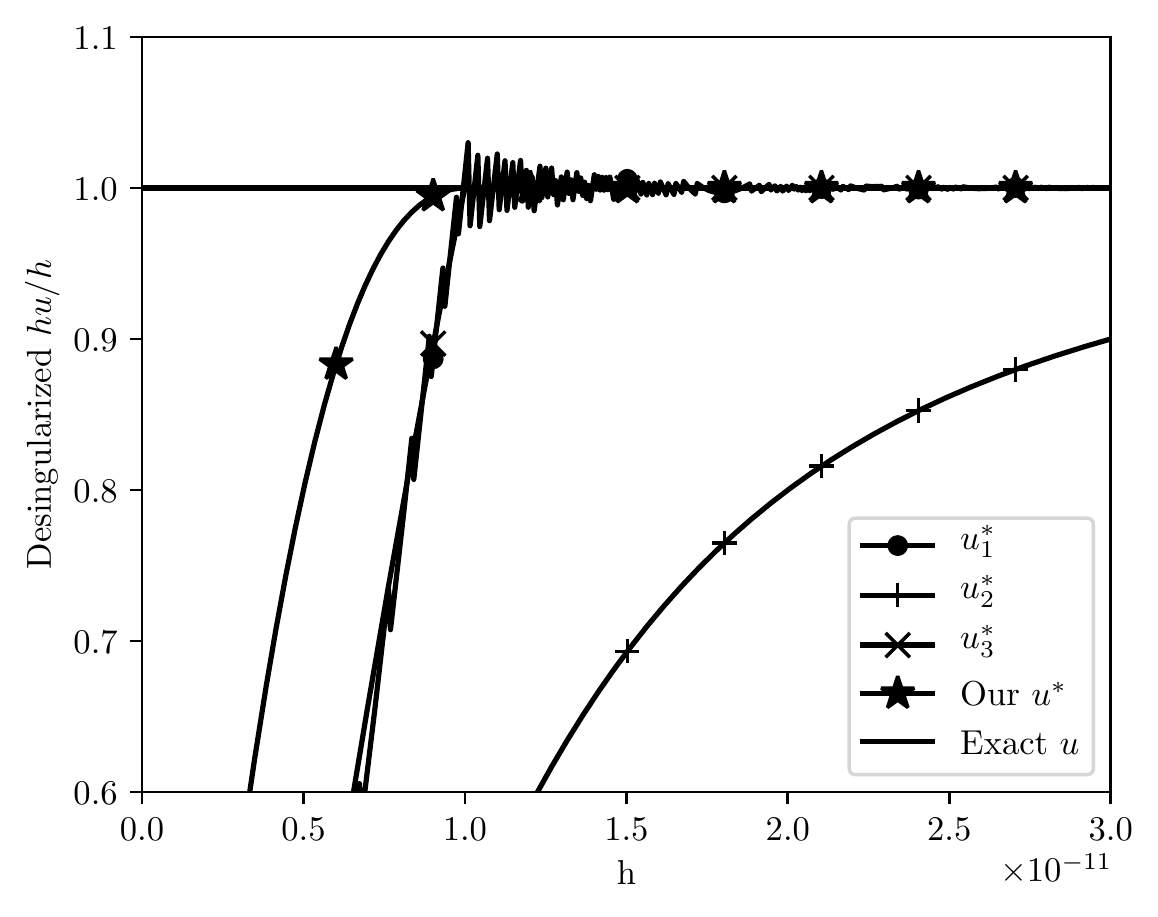}}
    \caption{Desingularization of the quantity $u=h/hu$ using existing and our proposed method with $\kappa=1.0 \cdot 10^{-11}$ for all cases. We compute the value of $hu/h$ using the different approaches for $u=1$ and different values of $h$. Notice how the single precision floating-point errors are clearly visible for $u_1^*$, that $u_2^*$ has a distinct kink, and that $u_3^*$ significantly underestimates the true magnitude of $u$.}
    \label{fig:desing}
\end{figure}

When the water depth becomes close to zero, the calculation of the particle velocity, $u = hu/h$ can become severely ill conditioned as the expression is prone to large round-off errors. To counter this, Kurganov~\cite{kurganov_2018} and Kurganov and Petrova~\cite{kp07} have proposed to \emph{desingularize} the term using expressions such as 
\begin{align}
    u^*_1 &= \frac{\sqrt{2}h(hu)}{\sqrt{h^4+\max(h^4, \kappa^4)}}
    \label{eq:desing_kp1}\\
    u^*_2 &= \frac{h\cdot hu}{h^2+\kappa^2},
    \label{eq:desing_kp2}
    \qquad\text{or}\\
    u^*_3 &= \frac{2h\cdot hu}{h^2+\max(h^2, \kappa^2)}.
    \label{eq:desing_kp3}
\end{align}
The suggestion of using $\kappa = \max(\Delta x^4, \Delta y^4)$ has some obvious problems for real-world simulations with $\Delta x > 1$, as pointed out in~\cite{sw12}. 
Furthermore, the desingularization of $u$ here is not very well-behaved.
Instead, we propose to use the following formulation 
\begin{equation}
    \begin{split}
    h^* &= \text{sign}(h)\max\Bigl(|h|, \min(h^2 / (2\kappa) + \kappa/2, \kappa)\Bigr)\\
    u^* &= hu/h^*.
    \end{split}
    \label{eq:desing_ours}
\end{equation}
The advantages of this formulation is that (i) $\kappa$ now controls directly what magnitudes of $h$ to desingularize, (ii) it improves the numerical stability as it avoids the square root and the problematic $h^4$ term, (iii) it yields a smooth transition between normal and desingularized values, and (iv) the effective magnitude of $h^*$ is controlled. 
Figure~\ref{fig:desing} shows the three desingularization strategies and clearly indicates that the desingularized quantity $h$ is better behaved with our proposed approach.

\subsection{Global wall boundary conditions}
\label{sec:wall_bc}
We have implemented wall and periodic boundary conditions to be able to run test cases to assess the numerical correctness of our simulator. 
For reflective von Neumann-type wall boundary conditions, we use so-called ghost cells that mirror the ocean state across the boundary to enforce a zero flux out of the domain.
In the absence of Coriolis forces, wall boundary conditions at $x=0$ are constructed by setting the ghost cell values according to
\begin{equation}
		\eta(-x,y) = \eta(x,y), \qquad
		hu(-x,y) = -hu(x,y), \qquad
		hv(-x,y) =  hv(x,y).
	\label{eq:tradWallBCx}
\end{equation}
Na\"{i}ve implementation of these conditions with the CDKLM scheme leads to gravitational waves along the boundary.
Special care is therefore required to balance the Coriolis potential energies in the reconstruction of $h$, so that
\begin{equation}
    \begin{split}
   0 &= h^W_{0,j} - h^E_{-1,j} \\
     &= \left(\eta_{0,j}  + H_{-1/2,j} - \frac{\dx}{2g}\Big[ (K_x)_{0,j}  + f_{0,j} v_{0,j} \Big]  \right)
					    - \left( \eta_{-1,j} +H_{-1/2,j} + \frac{\dx}{2g}\Big[ (K_x)_{-1,j} + f_{-1,j} v_{-1,j}\Big] \right),
    \end{split}
    \label{eq:KxWallBCReq}
\end{equation}
using \eqref{eq:hE}.
Here, we see that $H$ cancels immediately, whereas $\eta$ cancels by the use of \eref{eq:tradWallBCx}.
To make the last terms cancel, we need to set $f_{-1,j}v_{-1,j} = -f_{0,j}v_{0,j}$, which
means that in addition to \eqref{eq:tradWallBCx}, we need to enforce 
\begin{equation}
    f(-x, y) = -f(x,y).
\end{equation}
Note also that this results in $(K_x) = 0$ across the boundary from \eqref{eq:Kbackwards}.

\subsection{Moving wet-dry boundary and land mask}

The original CDKLM scheme breaks down on land as the gravitational wave speed $\sqrt{gh}$ becomes imaginary with negative depths.
To support a moving wet-dry boundary, we use a similar approach as Kurganov and Petrova~\cite{kp07} by adjusting the slope of $h$ so that all the reconstructed point values become non-negative.
For example, if we get $h_{i,j}^E < 0$  from \eqref{eq:hE}, we adjust the slope to become
\begin{equation}
	(K_x)_{i,j} = -f_{i,j} v_{i,j} - \frac{2g}{\dx}\Big[\eta_{i,j} + H_{i+1/2,j} \Big],
	\label{eq:KxE_adjust}
\end{equation}
and thus forcing $h_{i,j}^E = 0$ instead.

The moving wet-dry boundary is required for simulating, e.g., tsunamis and other phenomena in which the run-up is important. 
The approach nonetheless violates the steady-state reconstruction, causing non-physical waves and often also small time step sizes. For static wet-dry boundaries, we have therefore implemented wall boundaries following a land mask, i.e., ensuring a zero flux for the shore-line. In our approach, we explicitly set the flux between two neighboring cells to zero if any one of them is masked. The mask is represented by a unique no-data value in the bathymetry, thereby having no increase in the memory footprint. 
It should be noted that both of these approaches reduce the reconstruction to first order along the wet-dry boundary.

\subsection{Wind forcing and bed friction source terms}

The original formulation of CDKLM includes the bed slope source term, $B(Q)$, and Coriolis force, $C(Q)$, and we extend the scheme to also include the bed shear stress, $S(Q)$, and wind forcing, $W(Q)$. 
Bed shear stress is discretized using a semi-implicit formulation in the strong stability preserving Runge-Kutta scheme, 
\begin{equation}
    \begin{split}
    Q_{i,j}^{*} &= \left(  Q^n_{i,j} + \Delta t M(Q^n)_{i,j} \right) / \left(1+\Delta t S(Q_{i,j}^n)\right)\\
    Q^{n+1}_{i, j} &= \tfrac{1}{2} \Bigl(Q^n_{i, j} + \left(Q^*_{i, j} + \Delta t M(Q^*)_{i, j}\right) \Bigr) / \Bigl(1+\tfrac{1}{2}\Delta t S(Q_{i,j}^*)\Bigr),
    \end{split}
    \label{eq:rk2_semiimp}
\end{equation}
instead of \eqref{eq:rk2}.
This essentially means that we apply half of the bed friction using $Q^n$, and half of the friction using the predictive step $Q^*$. 

The wind stress comes from the atmospheric wind transferring momentum to the water column. We take the approach of Large and Pond~\cite{1981LargePond} and use 
\begin{align}
    \vec{\tau} = \frac{\rho_a}{\rho_w} C_D \, |W_{10}| \, W_{10},
\end{align}
in which $\rho_a=1.225$ is the specific density of air, $\rho_w=1025$ is the specific density of sea water, $W_{10}$ is the wind speed at 10 meters, and $C_D$ is the drag coefficient computed as 
\begin{equation}
    C_D = 10^{-3}
    \begin{cases}
        1.2 & \text{if}\quad |W_{10}| < 11~\mathrm{m/s},\\
        0.49 + 0.065\,|W_{10}| & \text{if}\quad |W_{10}| \geq 11~\mathrm{m/s}.
    \end{cases}
\end{equation}

The NorKyst-800 model stores its output, including the wind forcing, every hour. Because the time step of our simulator is significantly smaller than one hour, we use linear interpolation in time to get the most accurate wind stress source term for each time step. 
Given the wind stresses $W_{10}^a$ at time $t_a$ and $W_{10}^b$ at time $t_b$, we compute the wind stress at time $t \in [t_a, t_b]$ as a convex combination of the two
\begin{equation}
        W_{10}^n = (1-s)W_{10}^a + sW_{10}^b, \qquad
        s = (t - t_a) / (t_b - t_a).
    \label{eq:windstress_interpol}
\end{equation}
In space, we use specialized GPU texture hardware for bilinear interpolation (see e.g.,~\cite{brodtkorb_etal_state_of_the_art}),
resulting in a trilinearly interpolated wind stress. This is highly efficient, as the GPU has dedicated hardware for bilinear spatial interpolation that also includes caching. 
An added benefit is that the grid size of the wind stress data does not need to match up with the underlying simulation grid.

\subsection{Nesting the model into NorKyst-800}
A common technique in operational oceanography is to nest high-resolution local models within lower-resolution models that cover a larger domain.
For instance, the NorKyst-800 model is nested within TOPAZ, which is a 12-16~km resolution HYCOM-based coupled ocean and sea ice model with data assimilation~\cite{topaz_2017_xie}.
In this work, we nest our simulations into the NorKyst-800 model by setting initial ocean state and boundary conditions appropriately.
We use the so-called flow relaxation scheme~\cite{davies_76}, in which we have a relaxation zone between the internal computational domain and the external boundary conditions.
In this region, we solve the shallow-water equations as normal to obtain an internal solution $Q^{\mathit{int}}_{i,j}$, and then
gradually nudge it towards an external solution $Q^{\mathit{ext}}$ using 
\begin{equation}
    \begin{split}
	Q_{i, j} &= (1-\alpha) Q^{\mathit{int}}_{i, j} + \alpha Q^{\mathit{ext}}, \qquad
	\alpha = 1 - \tanh(d_{i, j}/d_0).
    \end{split}
    \label{eq:frs}
\end{equation}
Here, $d_{i, j}$ is distance in number of cells to the external boundary, and
the parameter $d_0$ is typically set to 2 or 3 to control the fall-off of the $\tanh$ relaxation function.

The implementation of the boundary conditions follows the same ideas as for the wind stress, using textures for linear interpolation in time and space.
We use float4\footnote{Textures on GPUs are originally designed to hold the color of one pixel as the color channels red, green, blue, and alpha, hence float4.} textures to hold our physical variables, $\eta, hu, hv$, because float3 is not supported as a texture format. We furthermore only use two textures,
as we pack the north and south boundaries into one texture, and similarly for the east and west.

\subsection{Coriolis force}
\label{sec:coriolis}
The original CDKLM scheme requires that the grid is aligned with the cardinal directions, which significantly restricts the use of the scheme. 
For higher latitudes this becomes especially pronounced, as both the latitude and direction towards north will vary across the domain.
We have therefore extended the scheme to support both a spatially varying north-vector and a spatially varying latitude, which is required to be able to run realistic simulations for the areas covered by the NorKyst-800. 
To account for the varying north vector, we project the momentum onto the local north and east vectors,
\begin{align}
    \begin{split}
        hu^e &= \vec{e} \cdot [hu, hv]^T, \qquad \vec{e}=[\cos(\theta), -\sin(\theta)]\\
        hv^n &= \vec{n} \cdot [hu, hv]^T, \qquad \vec{n}=[\sin(\theta), \cos(\theta)],
    \end{split}
\end{align}
and use these vectors when computing the source term $C(Q)$ so that
\begin{align}
    C(Q) = f \cdot \left[
    \begin{array}{c}
        0\\
        \vec{x}\cdot [hv^n, -hu^e]^T\\
        \vec{y}\cdot [hv^n, -hu^e]^T
    \end{array}
    \right].
\end{align}
in which the vectors $\vec{x} = [\cos(\theta), \sin(\theta)]$ and $\vec{y} = [-\sin(\theta), \cos(\theta)]$ project the result back into the $(x,y)$-coordinate system. 
Note that we also need to use the same approach in the reconstruction of $h$ as discussed in Section~\ref{sec:reformulation} to maintain the rotational steady states.

Our spatially varying Coriolis parameter is computed from the latitude, $l_{i, j}$, of a cell
\begin{align}
    f_{i, j} = 2\omega\sin (l_{i,j}), \qquad \omega = 7.2921\cdot 10^{-5} \text{rad/s},
\end{align}
or using a beta-plane model,
\begin{align}
    f_{i,j} = f_{\mathit{ref}} + \beta\cdot \Big( \vec{n} \cdot \big[(x_i-x_{\mathit{ref}}), (y_j-y_{\mathit{ref}})\big]^T \Big).
    \label{eq:betaplane}
\end{align}
Here, $f$ is linearized around a reference point, $(x_{\mathit{ref}}, y_{\mathit{ref}})$, with a slope $\beta$ in the direction of $\vec{n}$. Setting $\beta$ to zero yields a constant Coriolis parameter throughout the domain.

\begin{figure}
    \centering
    \includegraphics[width=0.21\textwidth, trim=0cm 0cm 2.3cm 0cm, clip]{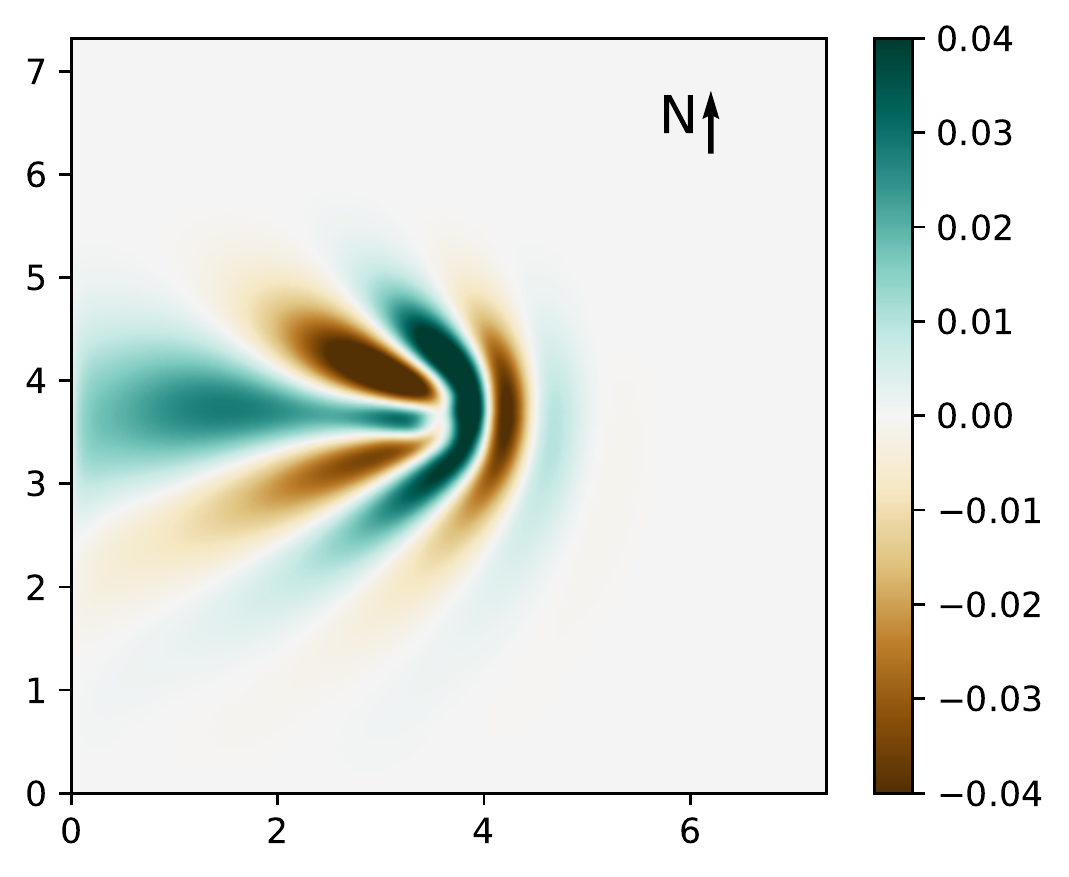}\quad
    \includegraphics[width=0.21\textwidth, trim=0cm 0cm 2.3cm 0cm, clip]{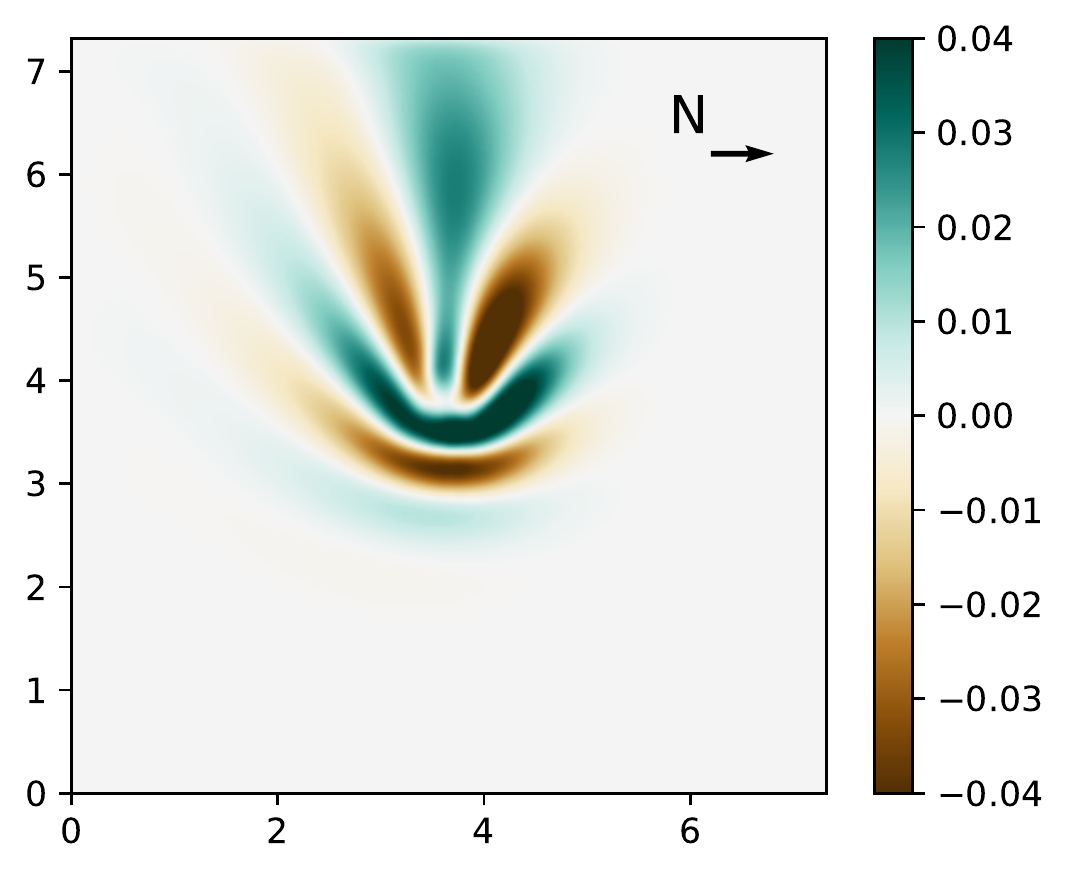}\quad
    \includegraphics[width=0.21\textwidth, trim=0cm 0cm 2.3cm 0cm, clip]{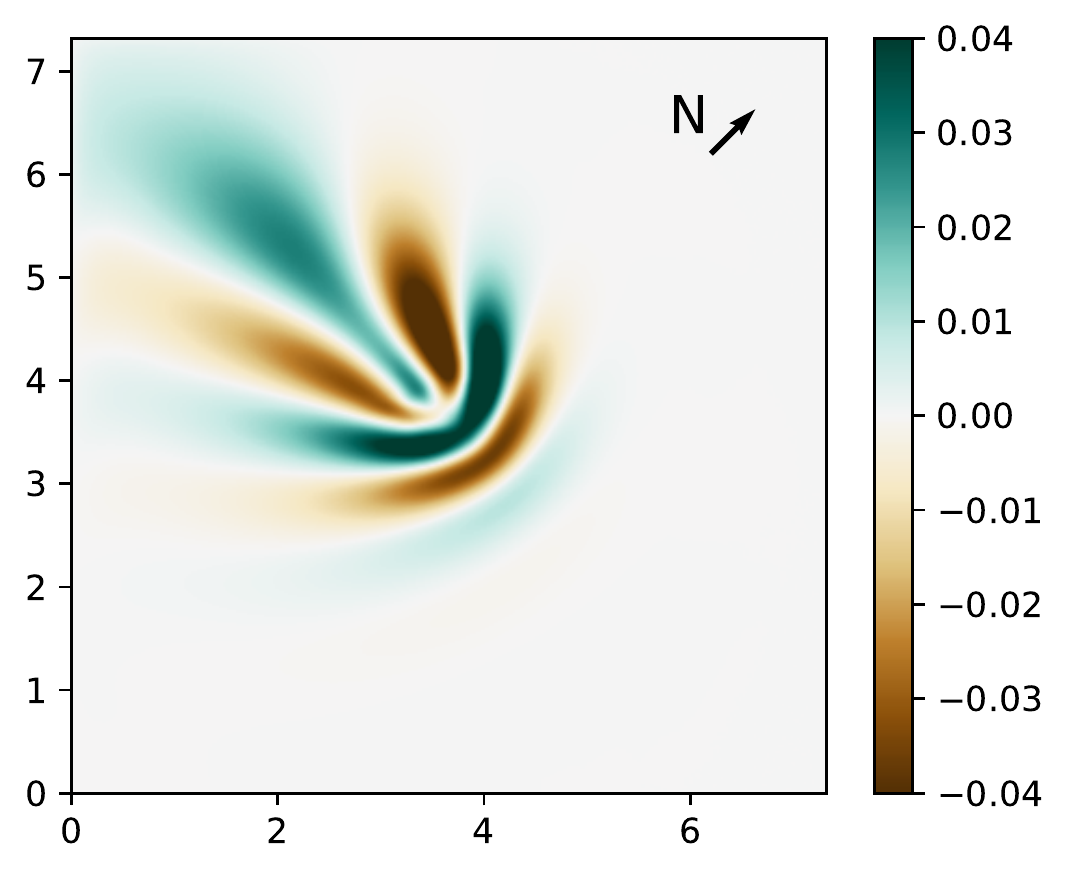}\quad
    \includegraphics[width=0.265\textwidth, trim=0cm 0cm 0cm 0cm, clip]{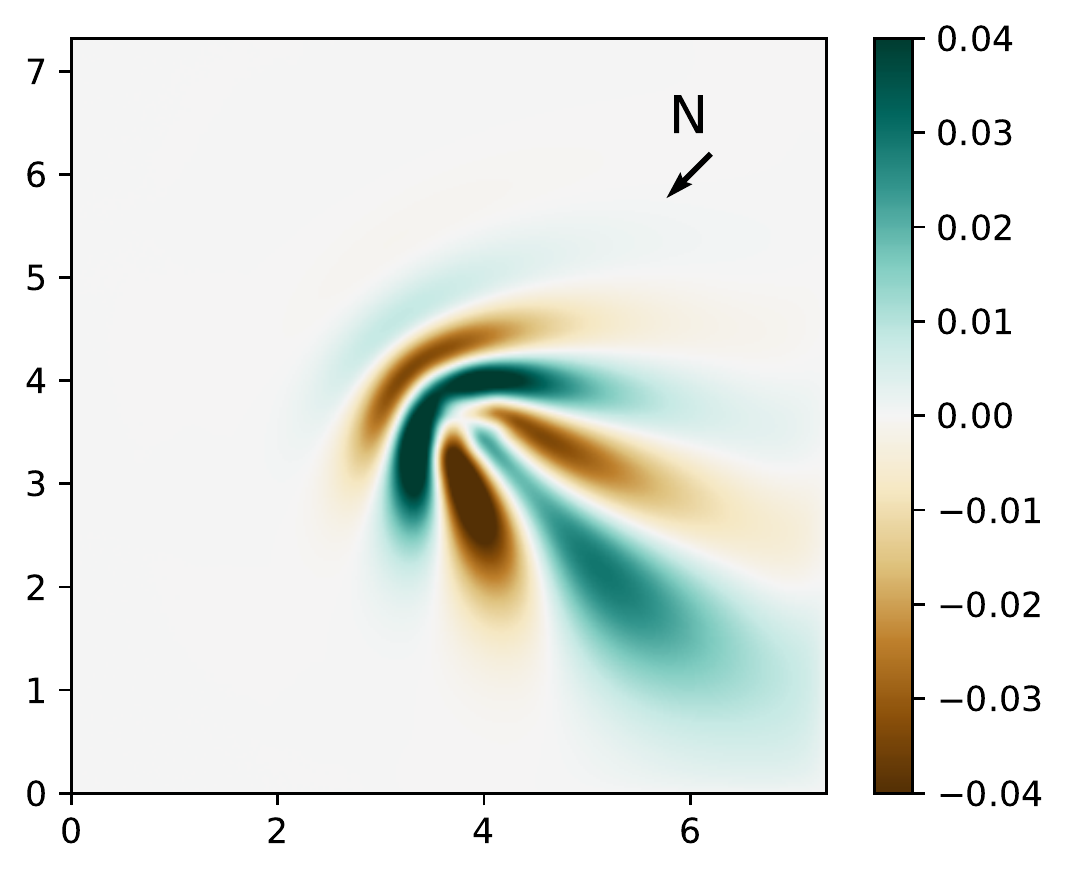}
    \caption{Planetary Rossby waves generated by a beta plane model for the Coriolis force and with different directions for north.
    The simulations are initialized by a rotating bump in the middle of the domain, which develops into Rossby waves that propagates westwards with wave energy slowly propagating eastwards.
    The different figures show that we get the same Rossby waves for arbitrary orientation of our domain.
    Axes are given in 1000~km, and the colormap shows $\eta$ in m.}
    \label{fig:varying_north}
\end{figure}

Figure~\ref{fig:varying_north} demonstrates how the angle to north plays an important role in the generation of planetary Rossby waves, caused by variations in the Coriolis force using \eqref{eq:betaplane}.
The figure shows the resulting $\eta$ after long simulations initialized with a rotating Gaussian bump in geostrophic balance in the middle of the domain, similar to the case presented in Holm et al.~\cite{gpuocean_testcases_preprint}.
Here, we have used a domain consisting of $350 \times 350$ cells with $\dx = \dy = 20\;\mathrm{km}$, a depth of $H = 50\;\mathrm{m}$, and the center of the domain corresponding to approximately $33^\circ$ north, with $f=8\cdot10^{-4}$ and $\beta=2\cdot10^{-11}$.
The simulation ends after approximately 35~days.
The figure clearly shows the effect of varying the north vector and hence the Coriolis source term.

\subsection{Bathymetry}
\label{sec:bathymetry}
The bathymetry in the NorKyst-800 model is given as cell midpoint values, and can not be used directly in our framework as CDKLM requires the bathymetry to be defined as a piecewise bilinear surface specified by point values at the cell intersections. 
It is still desirable that the averages of the resulting four intersection values end up to be equal to the NorKyst-800 cell average values, as large deviations can disturb the balanced relations in the provided ocean state.
Constructing such a piecewise bilinear surface from the cell averages is an ill-posed problem, and a simple averaging of cell values excessively smears important features such as deep fjords and straits. 

We have therefore devised a pragmatic iterative algorithm, which is based on reconstruction of slopes for each cell, similar to the reconstruction procedure in our numerical shallow-water scheme. Our initial guess is computed as follows:
\begin{enumerate}
    \item Compute the local gradient in each cell so that 
        \begin{align}
            [H_{i, j}^x, H_{i, j}^y] = \nabla H_{i,j}.
            \label{eqn:recon_1}
        \end{align}
    \item Evaluate the piecewise planar function at the four corners of each cell, 
        \begin{align}
            \begin{split}
                H_{i-1/2,j-1/2}^{i,j} &= H_{i, j} - \tfrac{1}{2} H_{i, j}^x - \tfrac{1}{2} H_{i, j}^y,\\
                H_{i+1/2,j-1/2}^{i,j} &= H_{i, j} + \tfrac{1}{2} H_{i, j}^x - \tfrac{1}{2} H_{i, j}^y,\\
                H_{i-1/2,j+1/2}^{i,j} &= H_{i, j} - \tfrac{1}{2} H_{i, j}^x + \tfrac{1}{2} H_{i, j}^y,\\
                H_{i+1/2,j+1/2}^{i,j} &= H_{i, j} + \tfrac{1}{2} H_{i, j}^x + \tfrac{1}{2} H_{i, j}^y.
            \end{split}
            \label{eqn:recon_2}
        \end{align}
    \item Average the estimate from the four cells meeting at one intersection point so that
        \begin{align}
            H^*_{i+1/2,j+1/2} = \tfrac{1}{4} \left( H_{i+1/2,j+1/2}^{i,j} + 
                                                    H_{i-1/2,j+1/2}^{i+1,j} + 
                                                    H_{i-1/2,j-1/2}^{i+1,j+1} + 
                                                    H_{i+1/2,j-1/2}^{i,j+1} \right)
            \label{eqn:recon_3}
        \end{align}
\end{enumerate}

\begin{figure}
    \centering
    \includegraphics[height=4.3cm, trim=0cm 0cm 1.5cm 0cm, clip]{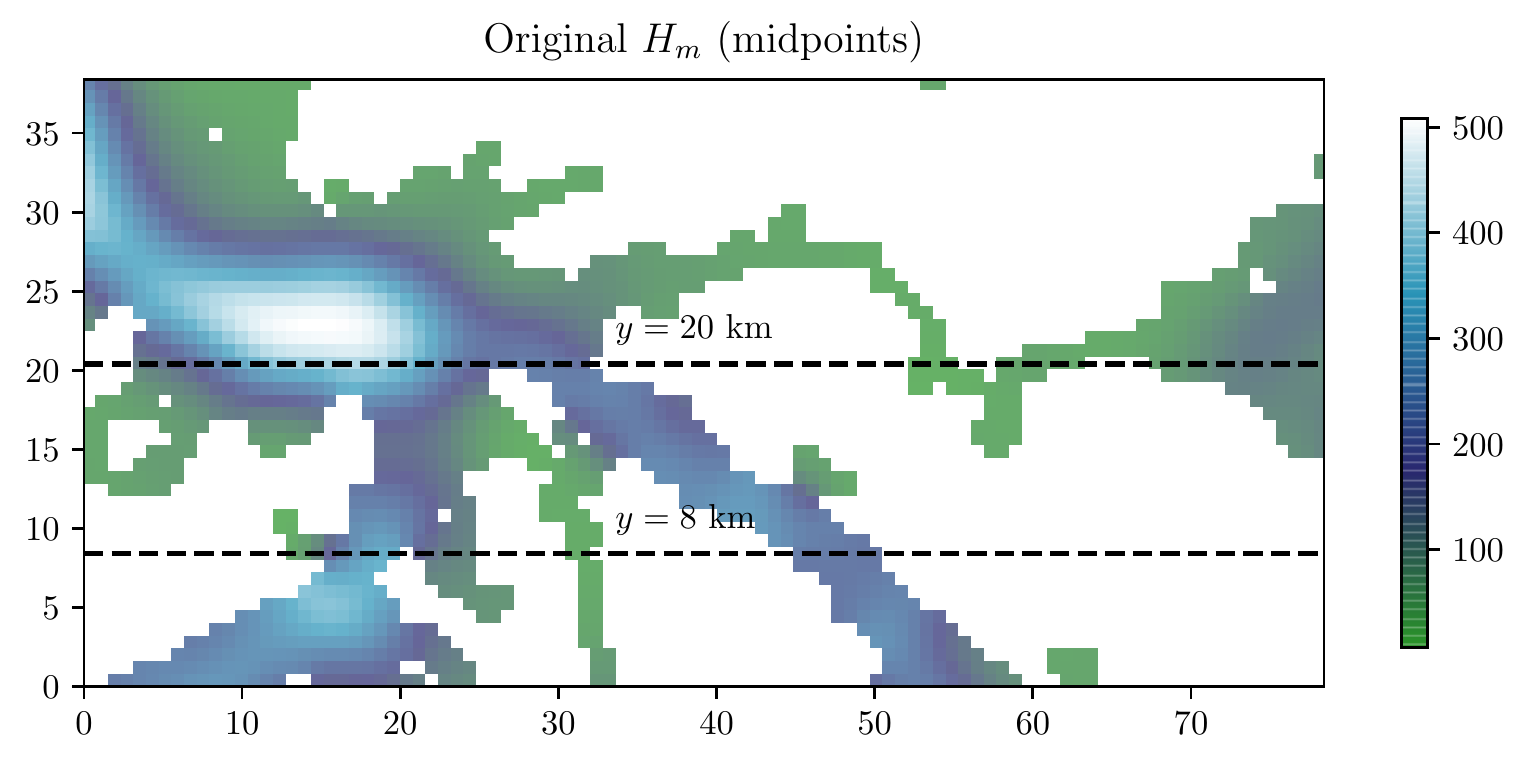}%
    \hfill%
    \includegraphics[height=4.3cm, trim=0cm 0cm 0cm 0cm, clip]{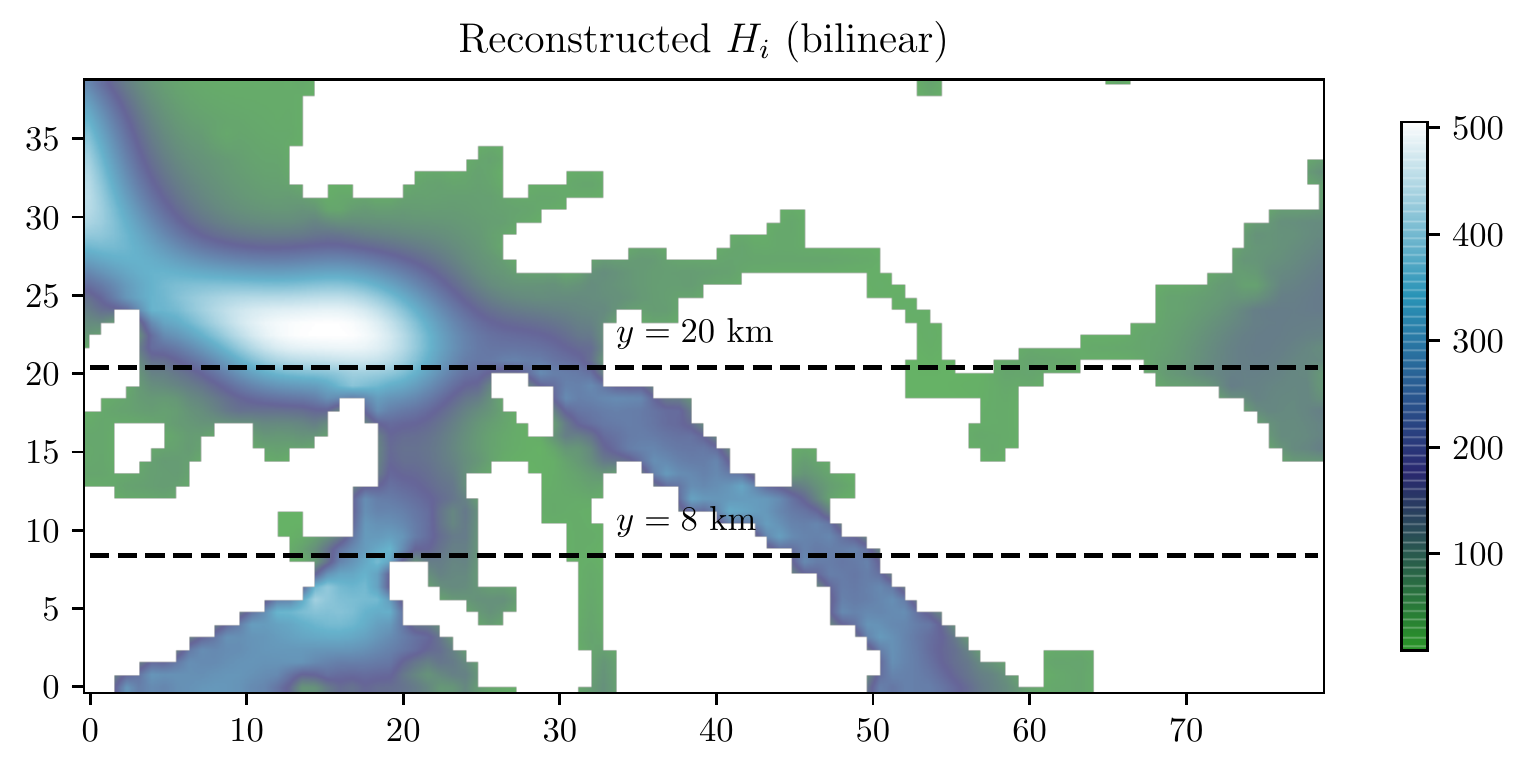}\\
    \includegraphics[height=5.5cm, trim=0cm 0cm 0cm 0cm, clip]{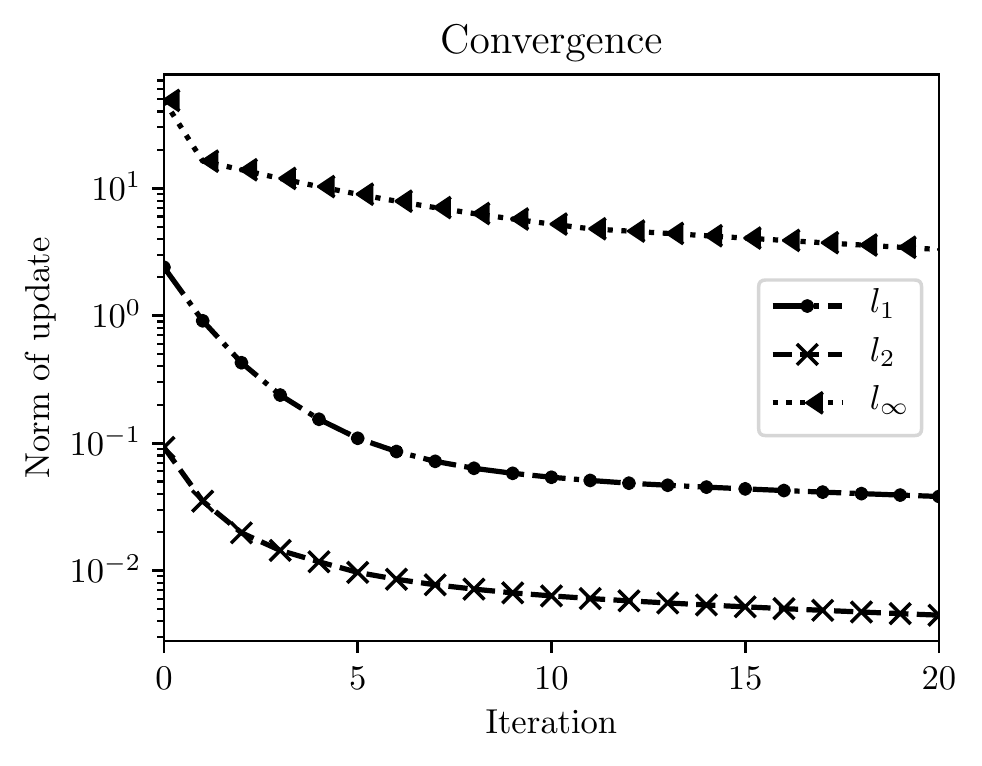}\hspace{2em}
    \includegraphics[height=5.3cm, trim=0cm 0cm 0cm 0cm, clip]{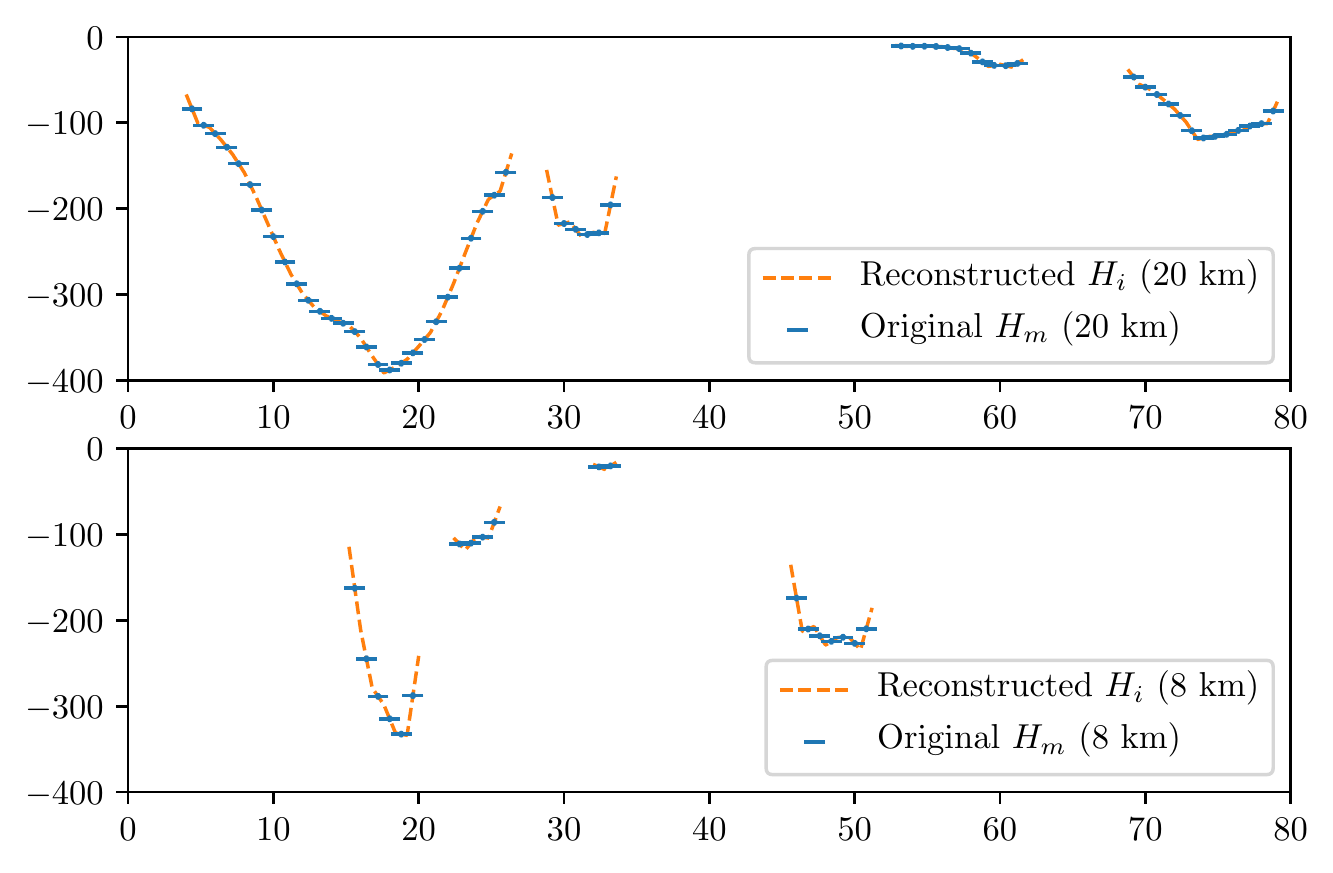}\hfill
    \caption{Reconstruction of the bathymetry at intersection points. Starting with the initial bathymetry $H_m$ defined at cell midpoints (upper left), we compute the $H_i$ values at intersections (upper right). The lower-left figure shows the convergence of our reconstruction algorithm, and the reconstructed bathymetry is plotted together with the original in the lower-right figure.}
    \label{fig:reconstruct_bathymetry}
\end{figure}

After obtaining the initial guess, we start an iterative procedure consisting of two stages. The first stage minimizes the difference between computed and true midpoint values, and the second dampens oscillations created by the first stage. We start by computing the difference between our estimate and the target bathymetry as 
\begin{align}
    \Delta H_{i, j} = H_{i, j} - H^*(i, j),
\end{align}
in which $H^*(i, j)$ means that we evaluate the piecewise bilinear estimate at midpoints. 
We then apply (\ref{eqn:recon_1})--(\ref{eqn:recon_3}) to compute an estimate of the difference at intersections, $\Delta H_{i+1/2, j+1/2}^*$, and subtract this difference from our initial guess, 
\begin{align}
    H^*_{i+1/2, j+1/2} := H^*_{i+1/2, j+1/2} + \Delta H_{i+1/2, j+1/2}^*.
\end{align}
This brings our updated midpoints of $H^*(i, j)$ closer to the target $H_{i, j}$, but it also results in unwanted oscillations, as there is nothing that limits the slopes of this piecewise bilinear surface. We therefore smooth $H^*$, and repeat the procedure until the update between two subsequent iterations is sufficiently small.
Around ten iterations is typically sufficient to create a piecewise bilinear surface that lies close to the target. Figure~\ref{fig:reconstruct_bathymetry} shows the result of our approach applied to the bathymetry from NorKyst-800 in the innermost parts of Vestfjorden (see also \secref{sec:case2_lofoten}).

\begin{figure}
    \centering
    \includegraphics[height=2.9cm,trim=0cm 0cm 12cm 1.2cm, clip]{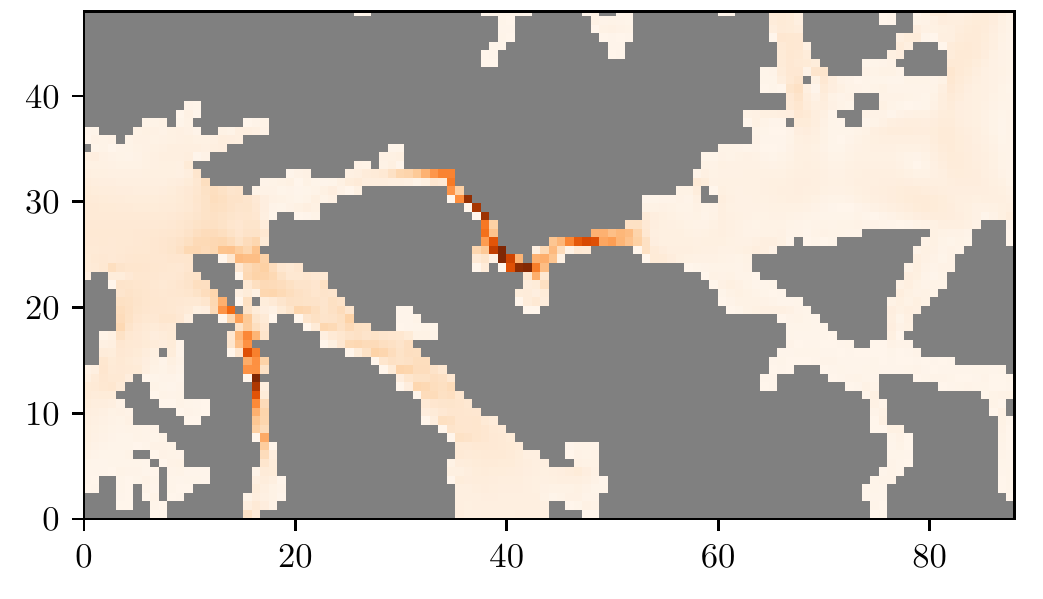}
    \hfill
    \includegraphics[height=2.9cm,trim=0cm 0cm 9.3cm 1.2cm, clip]{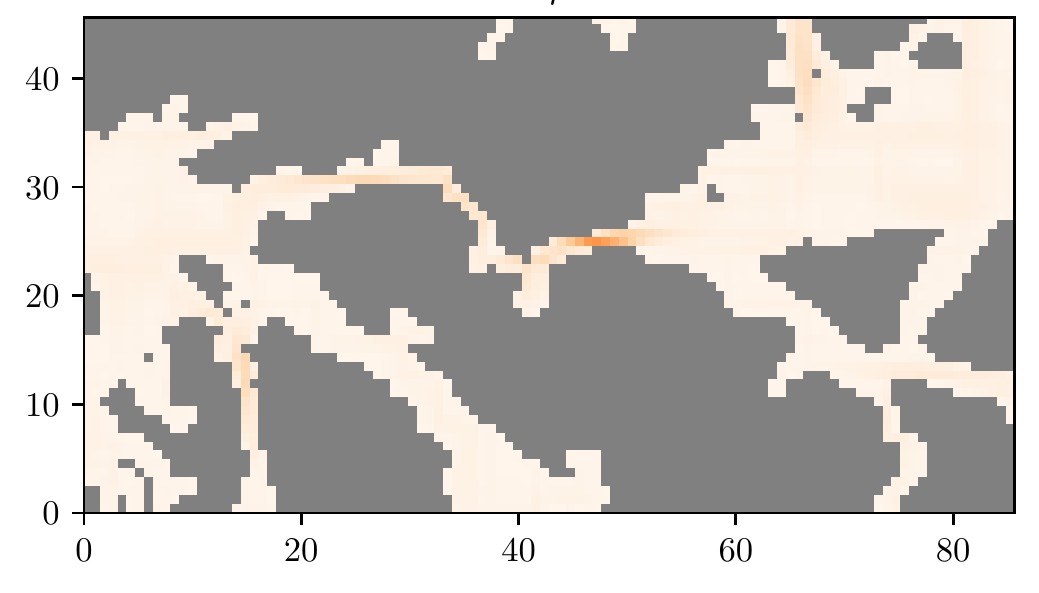}
    \hfill
    \includegraphics[height=2.9cm,trim=0cm 0cm 8cm 1.2cm, clip]{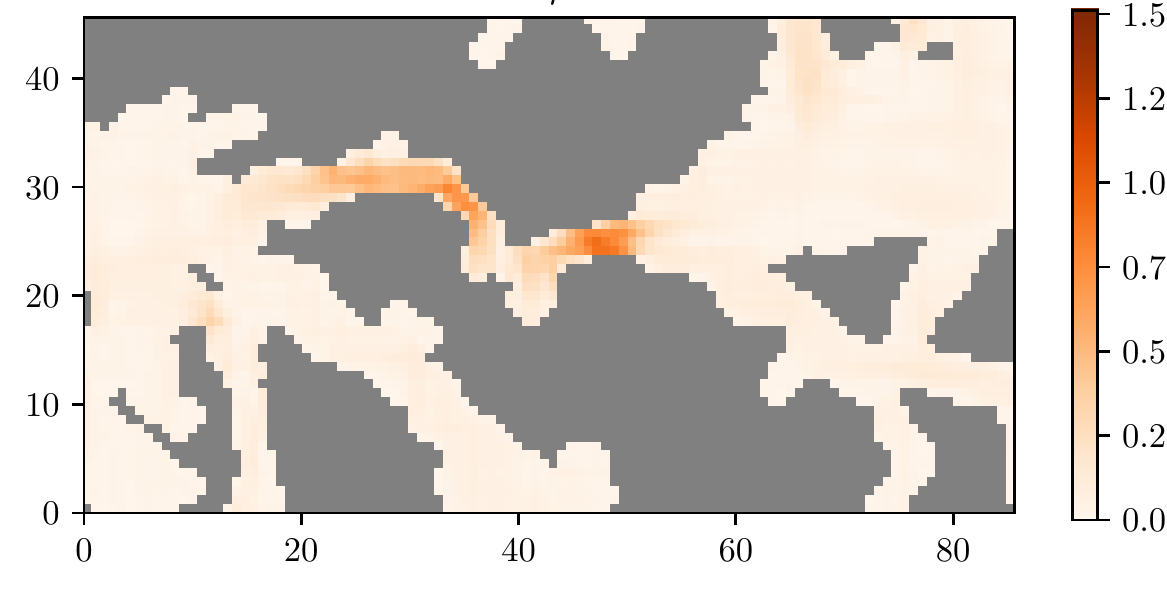}
    \caption{Widening of narrow straits, in which the land mask is eroded using binary erosion. The left figure shows particle velocities in the Tjeldsund strait in Lofoten/Vesterålen between the main land and Hinnøya (the bathymetry is shown in Figure~\ref{fig:reconstruct_bathymetry}). Left shows the NorKyst-800 reference, center shows original (reconstructed) bathymetry, and the right figure shows the result with one grid cell land erosion. The strait width and geometry limits the amount of water passing through it in the original bathymetry, but by eroding one cell of the land mask the amount of water is much closer to the reference.}
    \label{fig:erode_land}
\end{figure}

CDKLM is a full two-dimensional scheme, yet it exhibits a grid orientation bias close to the land mask. For example, we have experienced that the scheme has difficulties in propagating the momentum through narrow straits and canals that are only one or two cells wide. To ameliorate this, we can use binary erosion on the land mask to erode one row of cells at a time as shown in Figure~\ref{fig:erode_land}. We extrapolate $\eta$ into these areas using gray dilation (maximum of neighbors), and set the water depth equal the land value, which in the NorKyst-800 model is 5 meters. The process can be repeated to erode the land mask even more, and is required for tidal waves to properly propagate in features such as narrow fjords and straits. We have found that using single iteration is sufficient to propagate tides in and out of the fjords in the NorKyst-800 model.

\subsection{Refinement and coarsening of grid}
\label{sec:refinement}
The possibility to run simulations at different resolutions is attractive to quickly obtain low-resolution preliminary results, and also to resolve more of the dynamics for more detailed and accurate results with higher resolution.
We have included functionality in our simulation framework to increase and decrease the grid resolution with factors of two to support both of these scenarios.
We use a static grid refinement/coarsening performed on the initial conditions themselves, but the same approach can be used for efficient adaptive mesh refinement following the same approach as S\ae{}tra, Brodtkorb and Lie~\cite{saetra_brodtkorb_lie_2015}.

The physical variables are represented by cell average values, and refinement is based on a slope-limited reconstruction within each cell, which we evaluate at the refined grid cell centers.
As the bathymetry is specified by a piecewise bilinear surface, the high-resolution representation of the bathymetry evaluates the surface on the refined cell intersections.
Low-resolution representation of the physical variables on a coarsened grid is simply the average of the original cell values, and intersection values for the bathymetry are sampled from the corresponding original intersection values.
With a factor two coarsening, both these operations are well-defined.

Note that the boundary conditions and wind forcing terms are independent of the mesh resolution due to our use of GPU textures.

\section{Performance and accuracy of GPU implementation}
\label{sec:performance_evaluation}
We have previously evaluated the standard CDKLM scheme on several test cases which target oceanographic dynamics relevant for the shallow-water equations when implemented on a Cartesian grid~\cite{gpuocean_testcases_preprint}. 
In this work, however, we have reformulated the scheme in terms of $\eta$ instead of $h$, added several new source terms, and enabled varying latitude and north vector. It is therefore important to revisit the grid convergence test. 
We also measure computational performance of our simulation framework through assessing its weak scaling.

\subsection{Convergence of the numerical scheme} 
\begin{figure}
    \centering
    \hfill
    \begin{minipage}[c]{0.55\textwidth}
    \includegraphics[width=\textwidth]{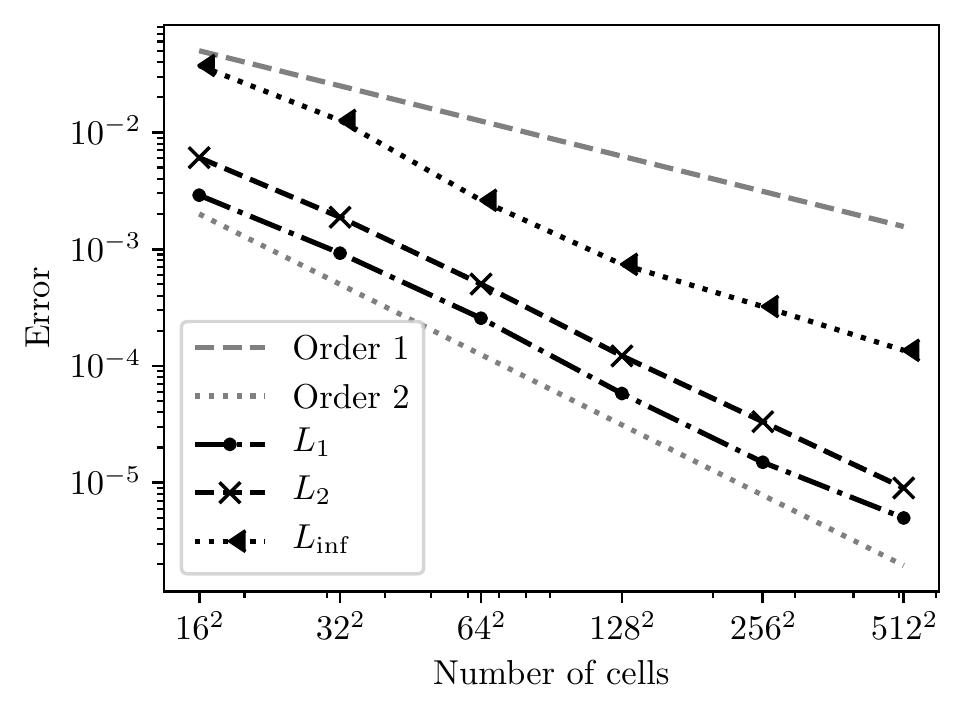}
    \end{minipage}\hfill
    \begin{minipage}[c]{0.35\textwidth}
    \includegraphics[width=\textwidth, trim=2.0cm 1.0cm 2.0cm 1.0cm, clip]{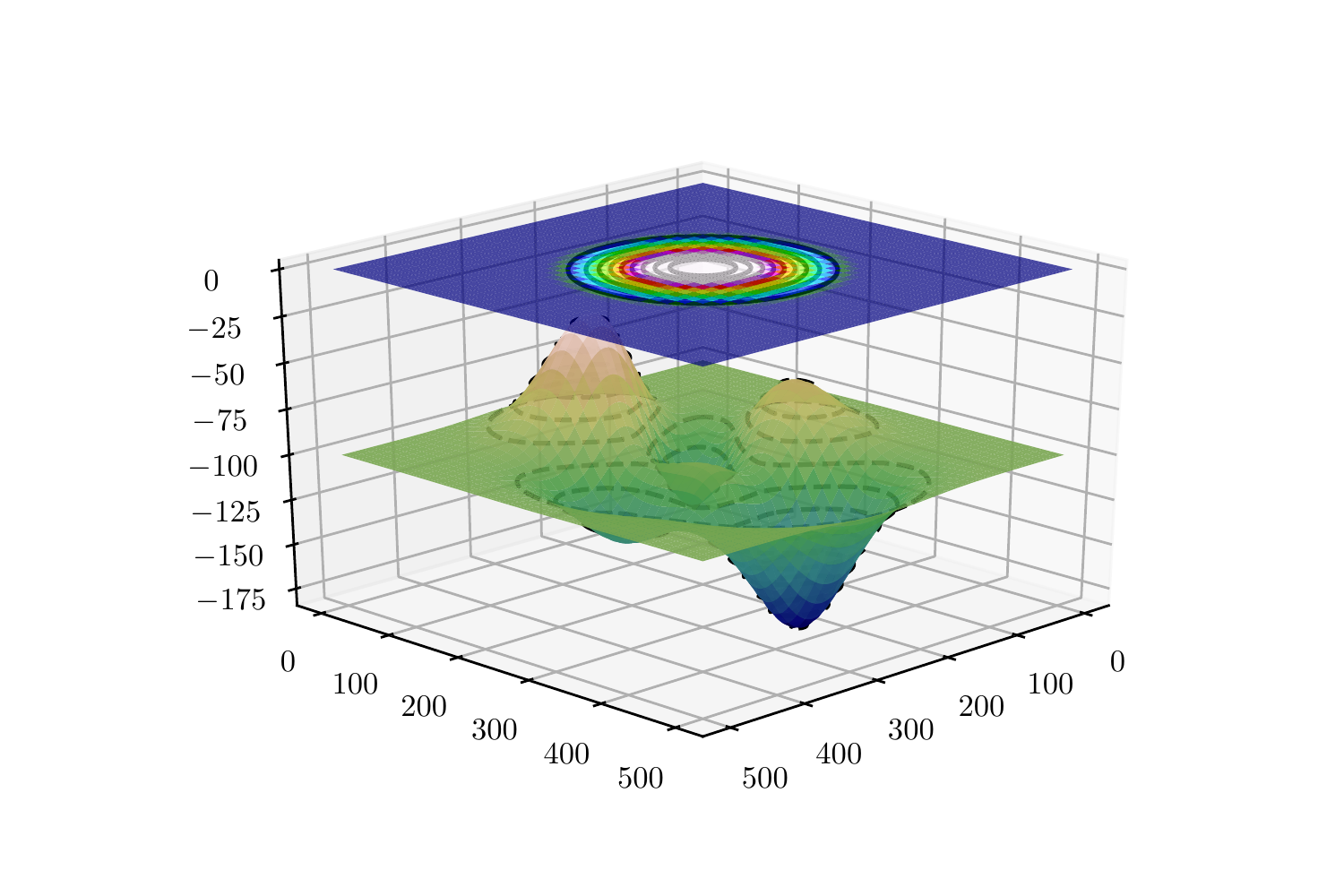}\\
    \includegraphics[width=\textwidth, trim=2.0cm 1.0cm 2.0cm 1.0cm, clip]{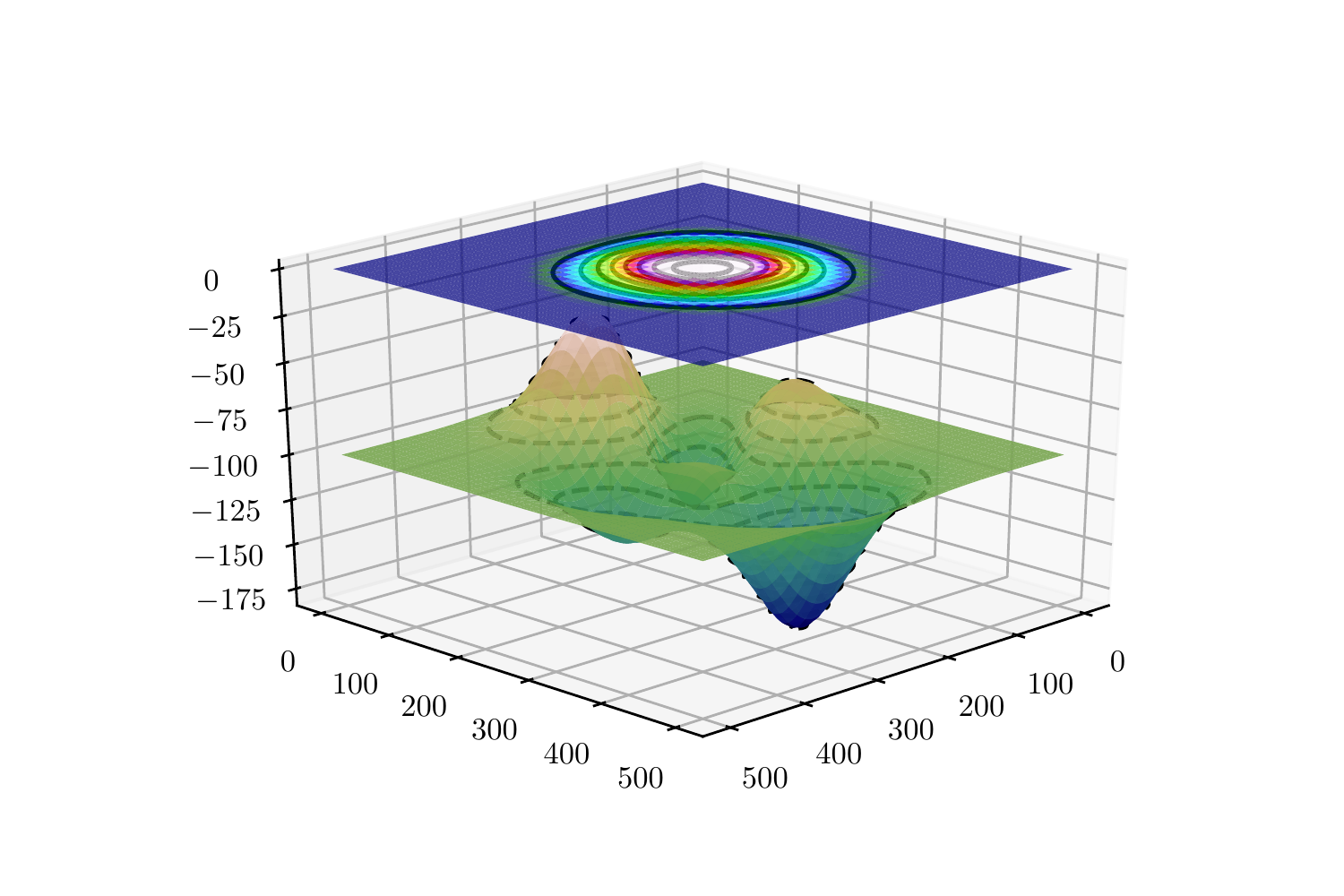}
    \end{minipage}
    \hfill\null
    \caption{Convergence plot of our numerical scheme. The numerical case has a water disturbance consisting of a cosine bump at the center of the domain, and the bathymetry consists of the smooth Matlab peaks function. The Coriolis force varies spatially across the domain according to variations in latitude and direction towards north. The left figure shows the self convergence of the numerical error compared to the reference solution with $1024^2$ cells. To the right we show the bathymetry with the sea-surface level initially (top) and at the end of the simulation (bottom).}
    \label{fig:convergence}
\end{figure}

To evaluate numerical convergence, we define a benchmark case covering $500~\mathrm{km} \times 500~\mathrm{km}$. We model the bathymetry after the Matlab \texttt{peaks} function,
\begin{align}
    \text{peaks}(s, t) = 3 (1-s)^2 e^{-s^2 - (t+1)^2} - 10 (s/5-s^3-t^5) e^{-s^2-t^2}-1/3 e^{-(s+1)^2-t^2}, \qquad s, t \in [-3, 3]
\end{align}
which yields a non-symmetric smooth surface.
We then define the equilibrium ocean depth as
\begin{equation}
    H(x,y) = 100 + 10 \cdot \mathrm{peaks}\left(\frac{6 x}{500} - 3, \frac{6y}{500} - 3 \right),
\end{equation}
with $x$ and $y$ given in km, so that the depth varies between 34 and 181 meters (see Figure~\ref{fig:convergence}).
We initialize the sea-surface level with a bump centered in the domain, 
\begin{align}
	\eta(r) = \begin{cases}
	    \tfrac{1}{2} \left( 1 + \cos\left(\tfrac{r}{c} \pi \right) \right), \quad &\mathrm{if} \; r \leq c, \\
	    0, &\mathrm{if} \; r  > c,
	\end{cases}
\end{align}
in which $r = \sqrt{(x-250)^2 + (y-250)^2}$ is the distance from the center of the domain, and 
$c$ controls the size of the bump, set to 300~km in our case. 
We assume that the North pole is located at  $(3000~\mathrm{km}, 3000~\mathrm{km})$, which means that the angle between the $y$-axis and the north vector varies from $33.7^\circ$ to $56.3^\circ$ across the domain and that the latitude varies from $56.0^\circ$ to $67.4^\circ$ north. These parameters roughly correspond to the North Sea.
Including land inevitably yield only first-order accurate fluxes, and we have thus deliberately not included such areas in our grid convergence test. 

The simulation is run for 800 seconds, after which we compute the difference for the whole domain between the solution and a reference computed on a $1024\times1024$ grid. Figure~\ref{fig:convergence} and Table~\ref{tab:convergence} show that we achieve second-order accuracy in both the $L_1$ and the $L_2$ norm, whilst the convergence in $L_{\mathrm{inf}}$ is somewhat slower.

\begin{table*}
\caption{Error of computed solutions at different resolutions and numerical reduction order for each refinement. The time step is fixed throughout these simulations. }
\label{tab:convergence}
\small 
\begin{center}
\begin{tabularx}{0.7\linewidth}{Xcrrrrrr}
	\textbf{Resolution} & \textbf{$\Delta t$} & \textbf{$L_1$} & \textbf{Order} & \textbf{$L_2$} & \textbf{Order}  &  \textbf{$L_{\mathrm{inf}}$} & \textbf{Order}  \\ 
	$16^2$ &  $1600/16$ & 0.002898 & - & 0.006072 & - & 0.037367 & - \\
    $32^2$ &  $1600/32$ & 0.000923 & 1.65 & 0.001877 & 1.69 & 0.012651 & 1.56 \\
    $64^2$ &  $1600/64$ & 0.000256 & 1.85 & 0.000502 & 1.90 & 0.002621 & 2.27 \\
    $128^2$ & $1600/128$ & 0.000058 & 2.14 & 0.000122 & 2.04 & 0.000741 & 1.82 \\
    $256^2$ & $1600/256$ & 0.000015 & 1.96 & 0.000033 & 1.87 & 0.000322 & 1.20 \\
    $512^2$ & $1600/512$ & 0.000005 & 1.58 & 0.000009 & 1.89 & 0.000136 & 1.25 \\
\end{tabularx}
\end{center}
\end{table*}

\subsection{Computational performance}
\begin{figure}
    \centering
    \includegraphics[width=0.55\textwidth]{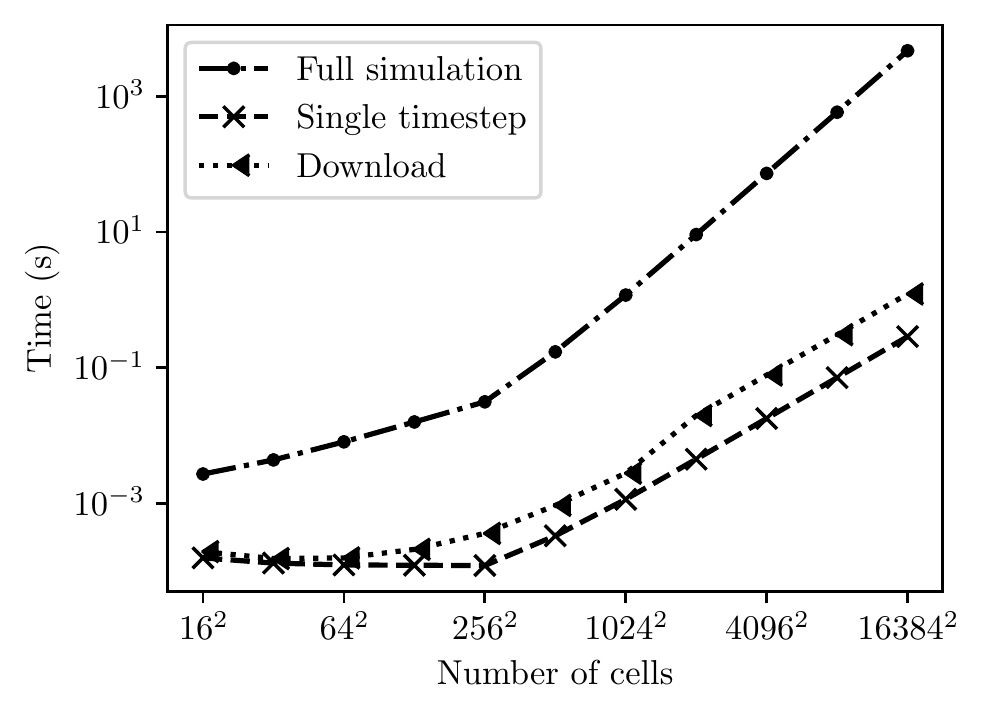}
    \caption{Weak scaling of our simulator. For small domain sizes, the runtime is dominated by overheads, but for domain sizes larger than approximately $512\times 512$, we see that the computational time scales with the number of cells. For each doubling of our domain, we perform approximately eight times as many operations (four times as many cells, and twice the number of time steps).}
    \label{fig:performance}
\end{figure}

To assess the computational performance, we continue to use the same benchmark, but run the simulation twice as long.
We also measure the cost of downloading the results from the GPU to the CPU.
Figure~\ref{fig:performance} shows the results for domains with $16^2$ to $16~384^2$ cells, running on a Tesla P100 GPU.
We see that for the larger domain sizes, we obtain perfect weak scaling, as a factor two grid refinement leads to a factor eight increase in computational complexity (the timestep size is cut in half due to the CFL condition, and we have four times the number of cells).
Cases with less than $512^2$ cells are dominated by overheads and hence, the cost of a single timestep is close to constant.
Time for downloading data scales linearly with the problem size, as is expected.
Again, we see the same behavior for small domains as for the computational performance.

\input{benchmark_results.tex}


\section{Real-world simulations}
\label{sec:real_world_sims}

Using the NorKyst-800 model system, the Norwegian Meteorological Institute issues daily ocean forecasts for the complete Norwegian coast and makes them publicly available through thredds servers. 
The results contain full 3D values as well as depth-averaged values, wind forcing, bathymetry, longitudes, latitudes, and angle towards north for each cell.
This means that it is simply a matter of accessing the file and cutting out the correct sections to be able to initialize our simulator.
To start a simulation within our framework, we simply need the full url to the specific forecast we want to nest our simulation within, and specify a desirable sub-domain.

\begin{figure}
    \centering
    \includegraphics[width=0.85\textwidth, trim=0cm 0.cm 0cm 0cm, clip]{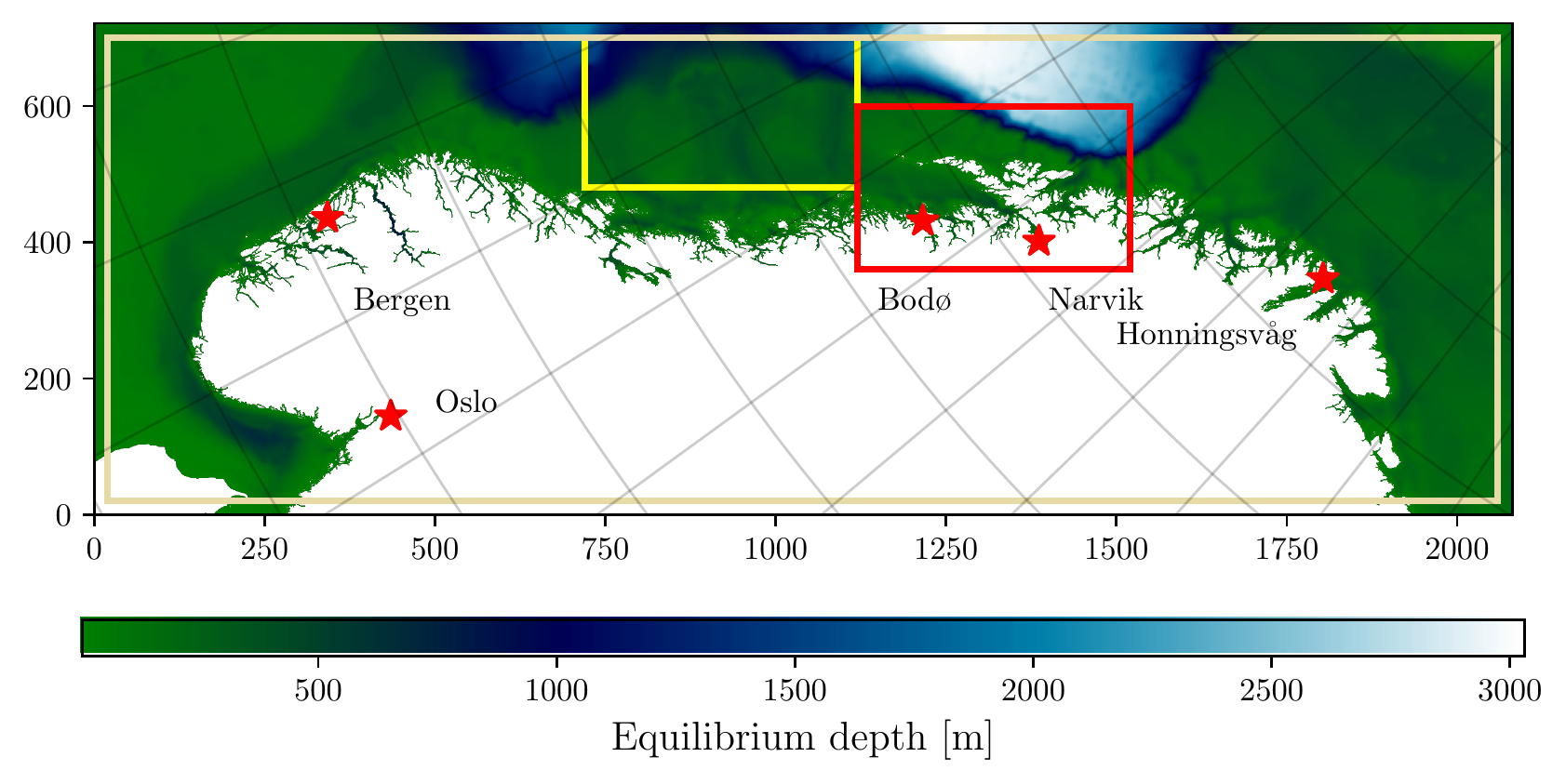}
    \caption{
    The domain covered by the NorKyst-800 model showing the equilibrium depth off shore.
    The location of Case 1 in the Norwegian Sea (yellow rectangle), Case 2 around the Lofoten archipelago (red rectangle), and Case 3 (beige rectangle) within the NorKyst-800 model domain, with red stars marking the locations for sea-level predictions.
    The values on the $x$- and $y$-axes are given in km, and the latitude-longitude grid shows how the direction towards north changes throughout the domain.}
    \label{fig:case_locations}
\end{figure}

We run our simulations on the three different sub-domains shown in Figure~\ref{fig:case_locations}.
Our first simulation is in the Norwegian Sea with no land values within the domain, and is therefore a fairly simple test for using real initial and boundary conditions.
The second case is centered on the Lofoten archipelago in Northern Norway, 
which is dominated by a complex coastline consisting of narrow fjords and a large number of islands, and serves as a challenging test for our reconstruction of the bathymetry and land mask.
Finally, we run almost the complete domain, leaving only 25 cells in each direction to serve as boundary conditions.
With the entire Norwegian coastline, it becomes important to account for the difference in the north vector throughout the domain (see the longitudinal lines and north arrows in Figure~\ref{fig:case_locations}), and we should see the tides following the coast as Kelvin waves.
The last case is also used to make sea-level predictions at five different locations, indicated as stars in Figure~\ref{fig:case_locations}.

All three cases are run with three different resolutions; a high resolution with a grid refined to 400~m cells, the original 800~m resolution of NorKyst-800, and a low resolution grid consisting of 1600~m cells. 
We use an adaptive time step, and update the time step every 20 minutes according to the CFL condition in \eref{eq:cfl} with a Courant number 0.8.
Each dataset provided from the NorKyst-800 forecasts covers 23 hours, which means that we only have boundary conditions to run simulations for the same time range.


\subsection{Case 1: Norwegian Sea}
The first case runs an open water domain in the Norwegian Sea and tests our ability to use the NorKyst data as initial and boundary conditions.
The sea-surface level changes significantly during the simulation time range as the tidal waves enter and exit the domain, and this dynamic is largely dependent on the boundary conditions.
The velocity field on the other hand, is dominated by already present features of slowly rotating dynamics, and will not change much during the simulation time range.
This enables us to test both that the initial conditions are correctly captured, and how well our simulator maintains complex rotating structures over time.

Figure~\ref{fig:norwegian_sea_23_hours} shows simulation results of the Norwegian Sea for all three different grid resolutions after 23 hours, compared to the reference solution from NorKyst-800.
First, we see that the contour lines for the sea-surface level in the left column show the same values at similar places for all simulations.
This shows that the boundary conditions allow tidal waves to propagate correctly in and out of the domain as intended.
The right column shows the corresponding particle velocities.
Note that the initial state for both the high-resolution and original runs are visually inseparable from the NorKyst-800 model state, whereas the low-resolution initial conditions slightly smear out the sharper features for the velocity. 
At 23 hours, we see that the high-resolution model to a large degree maintains the sharp features from NorKyst-800.
With the original resolution, however, the particle velocities resemble a slightly blurred image of the reference solution, and in the low-resolution simulation, the features are blurred even more. Also, we are starting to see clear signs of grid effects both internally and from the boundary.

\newcommand{\meanwidth}{0.85}
\newcommand{\meanwidthtop}{0.83}
\begin{figure*}[t!]
    \centering
    \subfloat{
        \includegraphics[width=\meanwidth\linewidth, trim=0 0.2cm 0 1.2cm, clip]{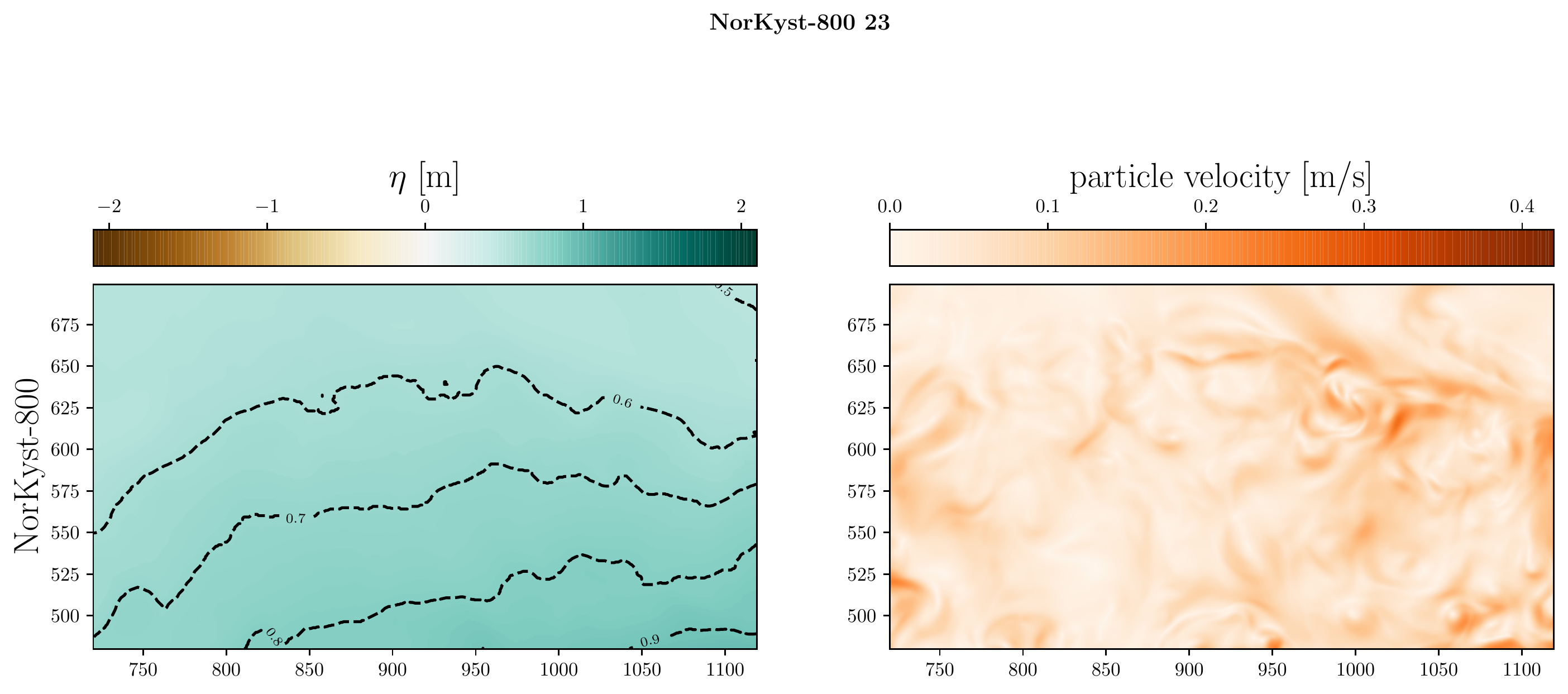}
    }
    \\
	\subfloat{
	    \includegraphics[width=\meanwidth\linewidth, trim=0 0.2cm 0 5cm, clip]{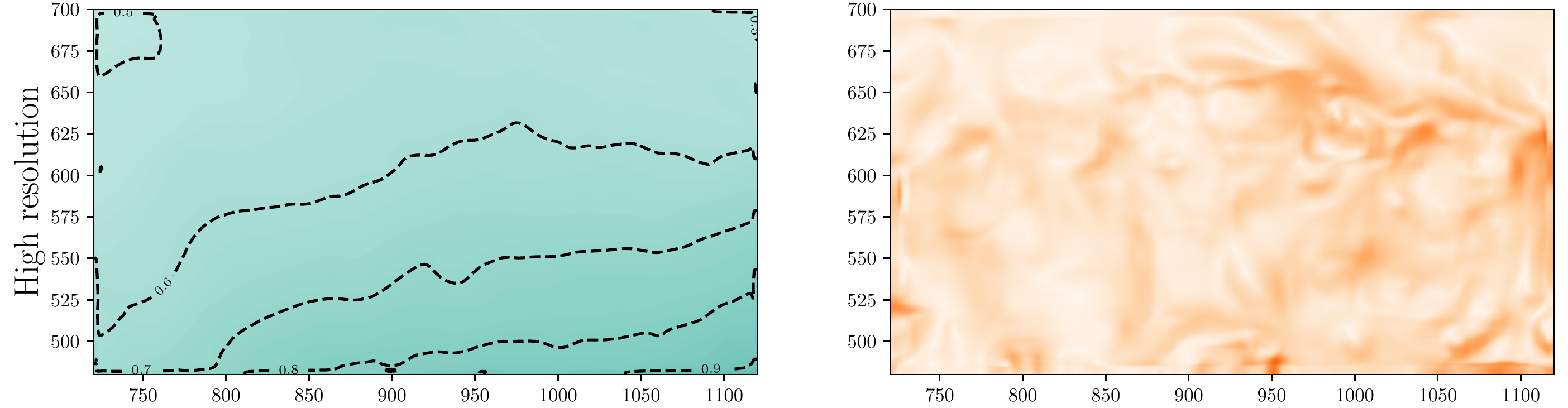}
	}
    \\
	\subfloat{
	    \includegraphics[width=\meanwidth\linewidth, trim=0 0.2cm 0 5cm, clip]{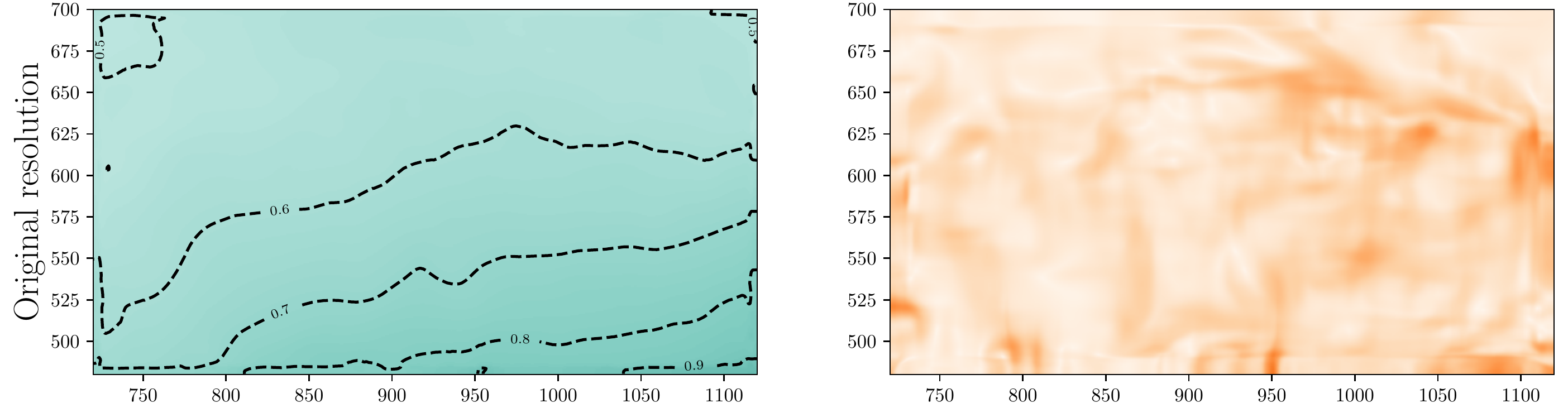}
	}
    \\
    \subfloat{
        \includegraphics[width=\meanwidth\linewidth, trim=0 0.2cm 0 5cm, clip]{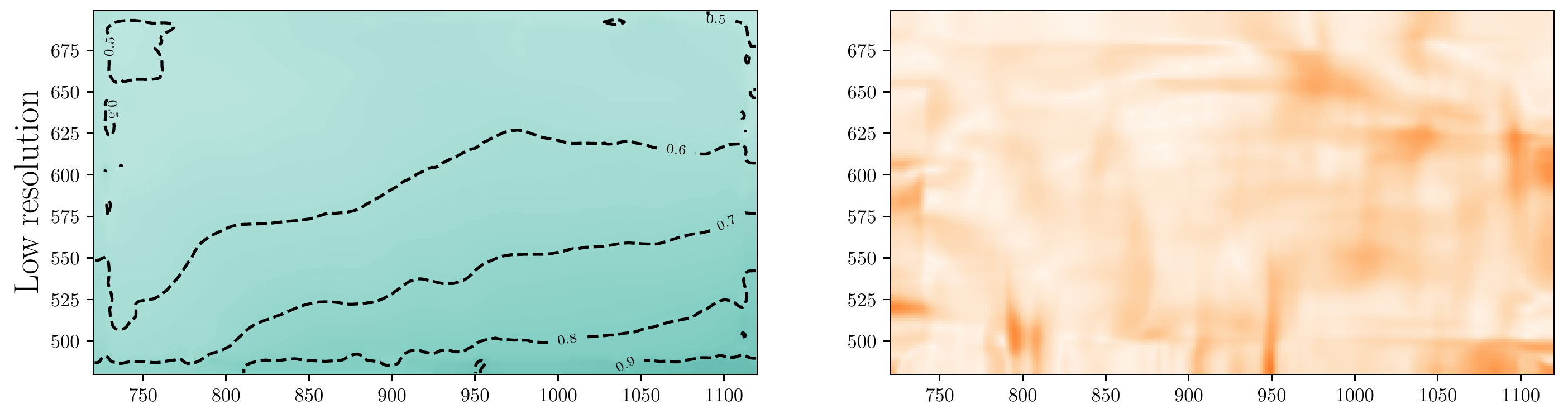}
    }
    \caption{Case 1 of the Norwegian Sea after 23 hours, comparing the reference solution from NorKyst-800 with our simulations using three different grid resolutions. All simulations obtain sea-surface level (left) similar to the reference solution, meaning that the boundary conditions allow the tides to correctly enter and exit our domain. From the particle velocities (right), we see that only the high-resolution model is able to maintain sharp features, whereas the low-resolution simulation has strong signs of grid effects. The values on the $x$- and $y$-axes are in km relative to the location in the complete domain.}
    \label{fig:norwegian_sea_23_hours}
\end{figure*}

\subsection{Case 2: Lofoten}
\label{sec:case2_lofoten}

\newcommand{\meanwidthLofoten}{0.83}
\begin{figure*}[t!]
    \centering
    \subfloat{
        \includegraphics[width=\meanwidthLofoten\linewidth, trim=0 0.2cm 0 1.2cm, clip]{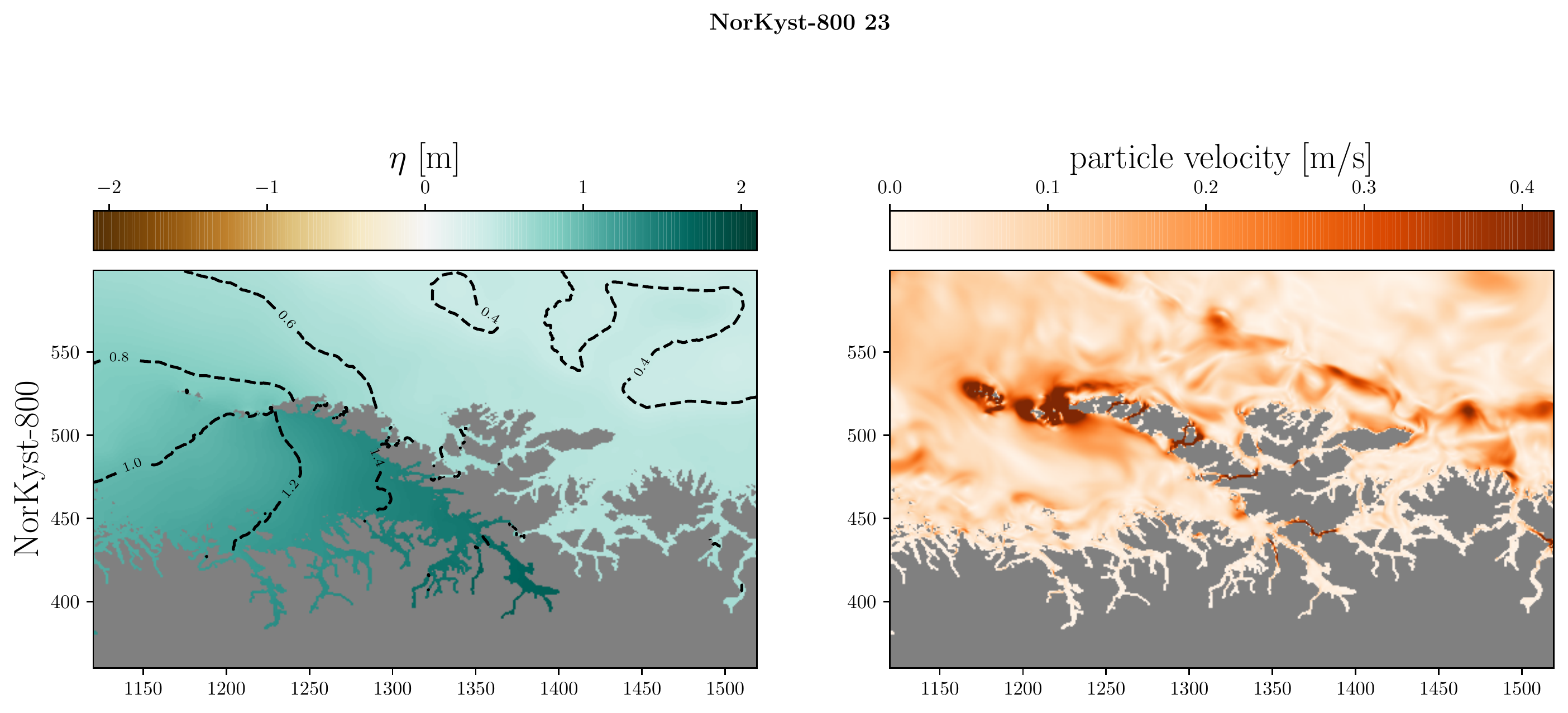}
    }
    \\
	\subfloat{
	    \includegraphics[width=\meanwidthLofoten\linewidth, trim=0 0.2cm 0 4.8cm, clip]{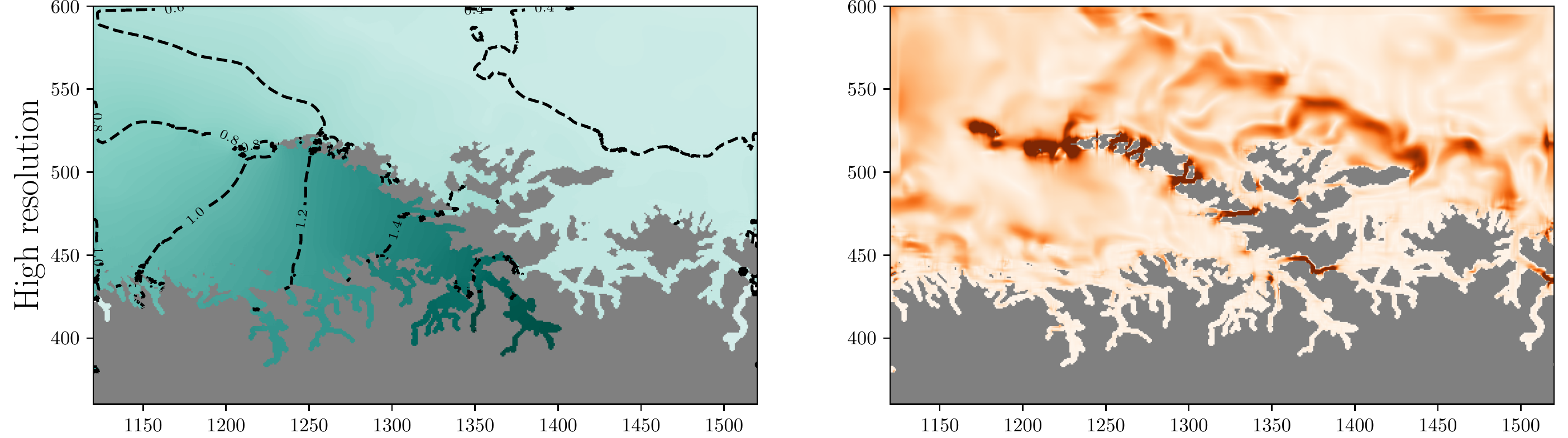}
	}
    \\
	\subfloat{
	    \includegraphics[width=\meanwidthLofoten\linewidth, trim=0 0.2cm 0 4.8cm, clip]{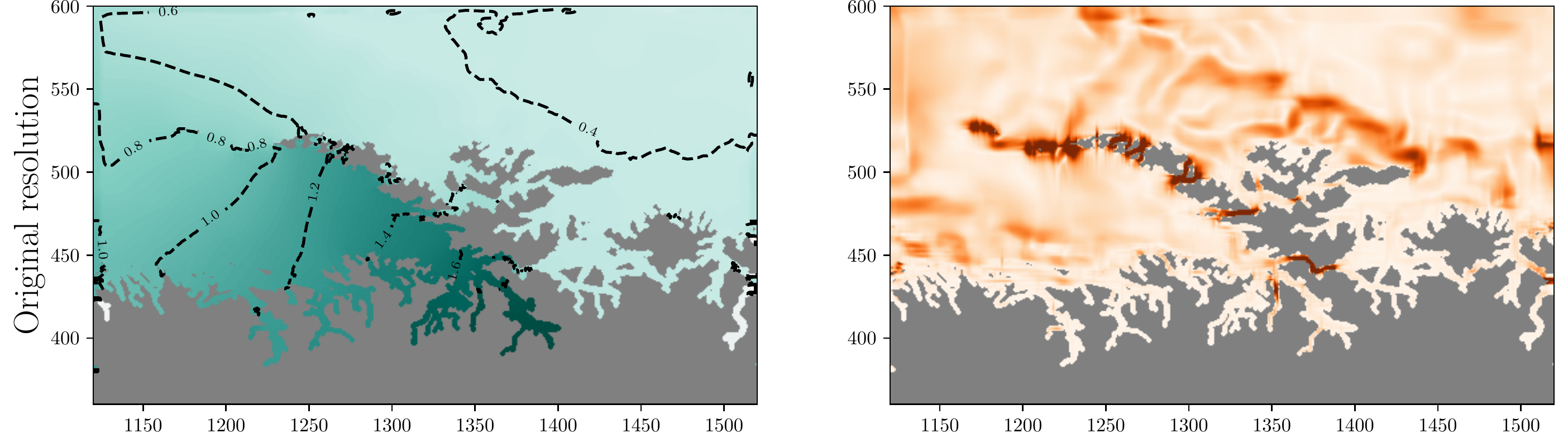}
	}
    \\
    \subfloat{
        \includegraphics[width=\meanwidthLofoten\linewidth, trim=0 0.2cm 0 4.8cm, clip]{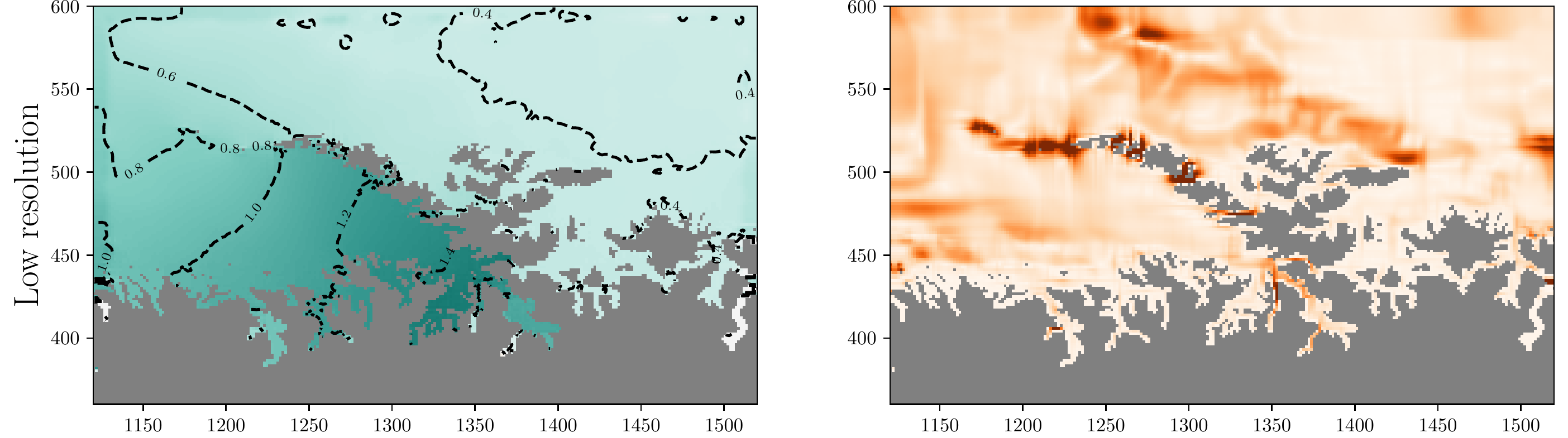}
    }
    \caption{Case 2 around the Lofoten archipelago after 23 hours, comparing the reference solution from NorKyst-800 with our simulations using three different grid resolutions. The values on the $x$- and $y$-axes are in km relative to the location in the complete domain. Our sea-surface levels (left column) are similar but slightly higher within Vestfjorden compared to the reference solution. From the particle velocities (right), we see that we get stronger currents than the reference. The level of details in our simulations are stronger with higher grid resolution, whereas the low-resolution simulation blurs the structures and are influenced by grid effects. 
    }
    \label{fig:lofoten_23_hours}
\end{figure*}

The second case is centered around the Lofoten archipelago in Northern Norway.
This region has a complex coast line consisting of a large number of fjords, straits and islands.
In addition to the aspects tested in the first case, this case tests the reconstruction of the bathymetry at intersections and the handling of dry zones through the complex land mask. 
We expect to see a Kelvin wave following the coast, but because the Lofoten archipelago protrudes into the Atlantic, some of the wave will be partially trapped and amplified in Vestfjorden between Lofoten and the main land.
Furthermore, as can be seen from the depth map in Figure~\ref{fig:case_locations}, the edge of the continental shelf crosses through the upper right corner of the sub-domain, and we can here see a line of eddies driven by baroclinic instabilities.

We again run simulations with the three different grid resolutions, and
Figure~\ref{fig:lofoten_23_hours} shows the final simulation state after 23 hours, with a high tide in Vestfjorden.
All three simulations show that the contour lines for the sea-surface level are slightly off compared to the reference solution, with marginally more water into the fjord.
Similarly as in Case 1, the low-resolution simulation is unable to preserve the fine structures in the currents and shows grid effects.
The original and high grid resolutions are also unable to preserve the exact eddy structure as found in the reference solution, but we can instead more clearly see a consistent coastal current with natural irregularities. 
This is consistent with what we can expect using a barotropic model.
These results also show a larger area with high particle velocity around the smallest islands in our simulations, possibly due to the reconstructed bathymetry, which widens some straits by the landmask erosion.

\subsection{Case 3: Norway}
Our third case consists of almost the complete domain covered by the NorKyst-800 model.
We leave out the 25 outermost cells and use these as boundary conditions.
With the complete domain, we cover a large enough area to see the Kelvin waves forming in the middle of the domain and travelling north.

\begin{figure*}[t!]
    \centering
    \subfloat{
        \includegraphics[width=0.98\linewidth, trim=0 0.2cm 0 1.2cm, clip]{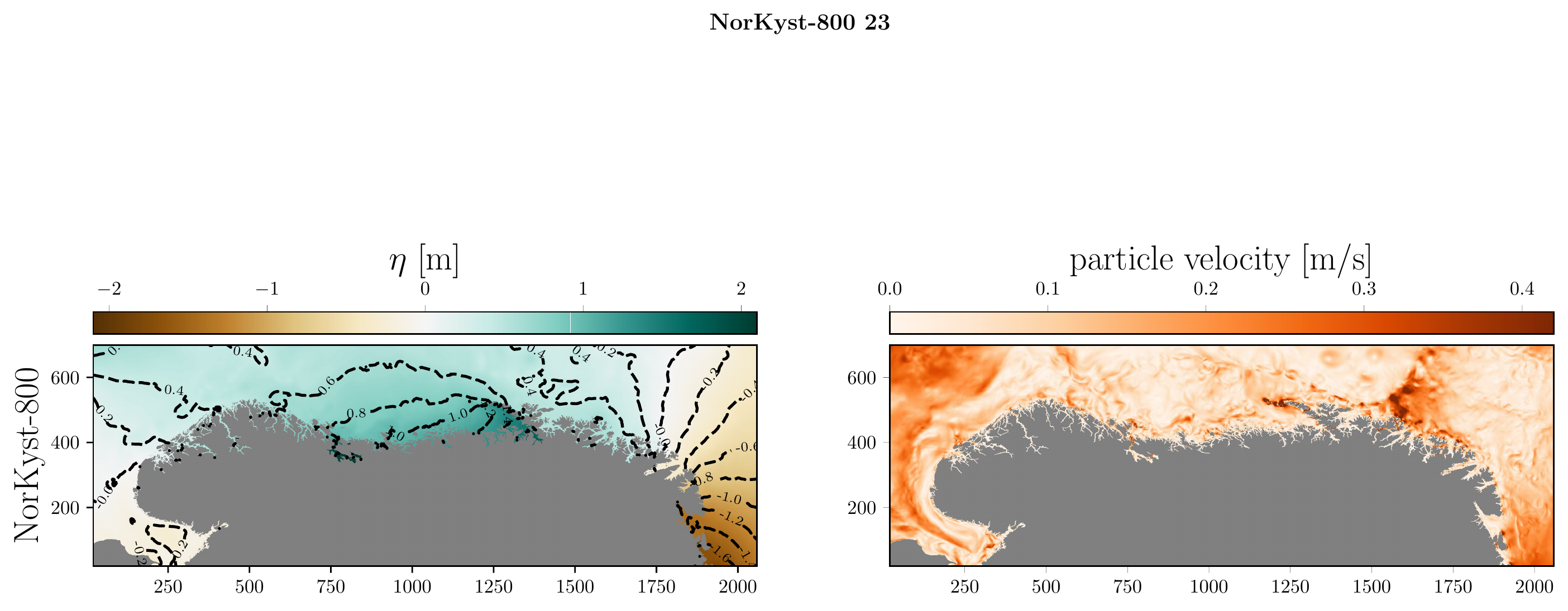}
    }
    \\
	\subfloat{
	    \includegraphics[width=0.98\linewidth, trim=0 0.2cm 0 6.2cm, clip]{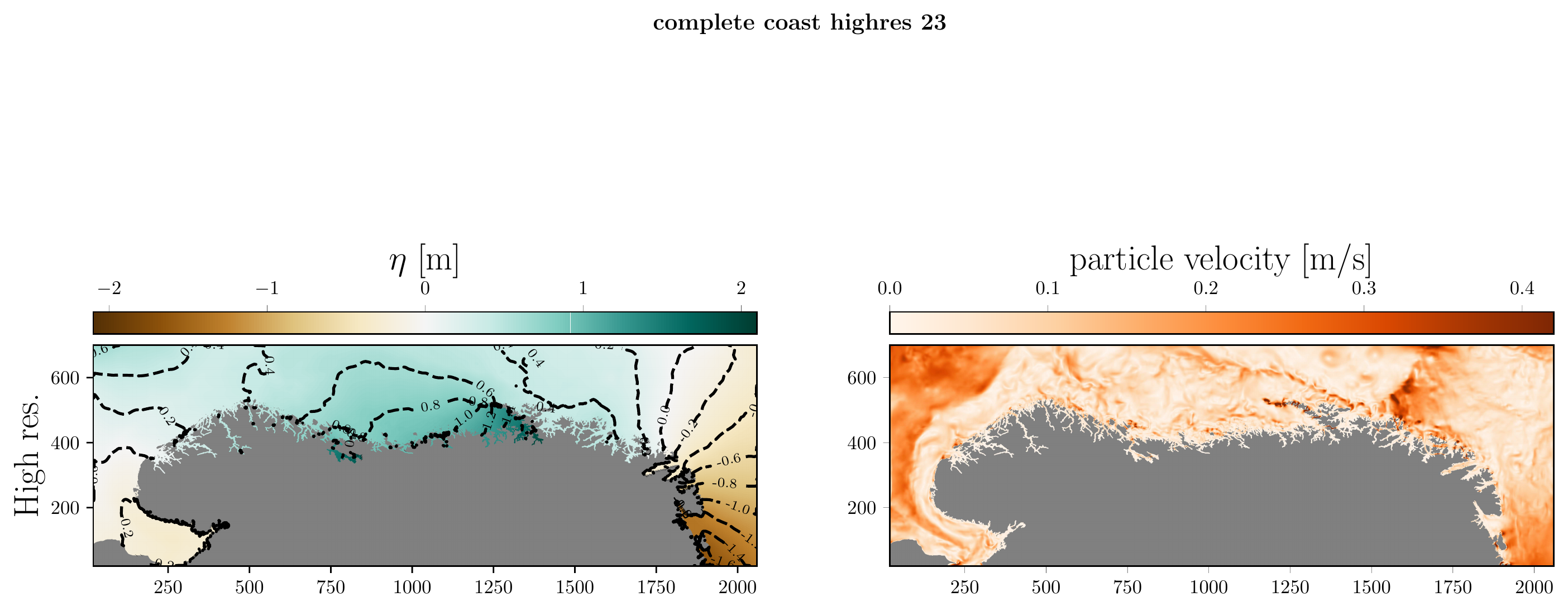}
	}
    \\
	\subfloat{
	    \includegraphics[width=0.98\linewidth, trim=0 0.2cm 0 6.2cm, clip]{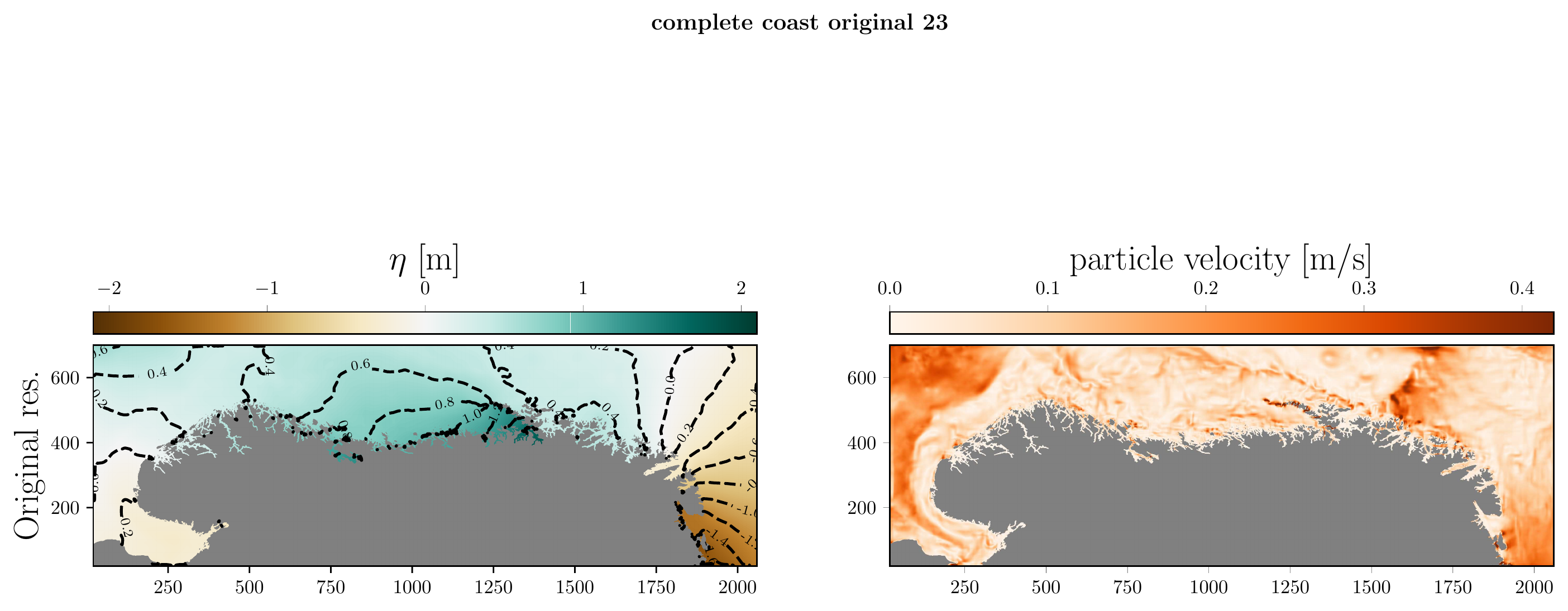}
	}
    \\
    \subfloat{
        \includegraphics[width=0.98\linewidth, trim=0 0.2cm 0 6.2cm, clip]{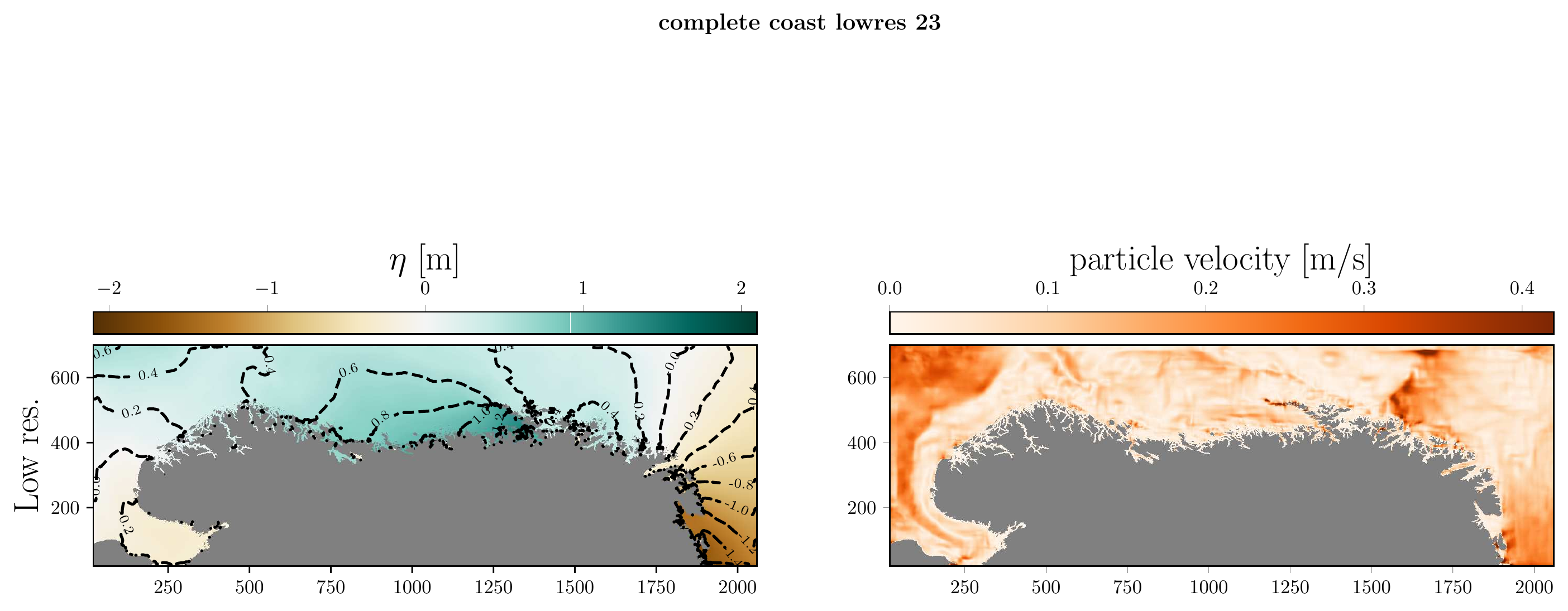}
    }
    \caption{Simulation result for Case~3 containing almost the complete domain used by NorKyst-800 after 23~hours.
    The top row shows the reference solution from NorKyst-800, followed by our simulation results using three different grid resolutions below. The values on the $x$- and $y$-axes are in km relative to the location in the complete domain.
    The sea-surface levels (left) show how the tide varies along the coast, with low tide in the Oslofjord (lower left corner), high tide in the middle of the domain, and low tide again towards the border with Russia (lower right corner).
    Our results are consistent with the reference solution for all grid resolutions.
    The particle velocities (right column) are qualitatively similar for the reference solution (NorKyst-800), and our simulations with high and original resolution.
    Even at this scale, it is apparent that the low resolution grid smears the features in the ocean currents.
    }
    \label{fig:complete_coast_23_hours}
\end{figure*}

Figure~\ref{fig:complete_coast_23_hours} shows the simulation results after 23 hours in the same manner as the earlier figures.
We now see from the sea-surface level (left column) that there is a low tide in the middle of the domain, whereas there are high tides both in the north at the Russian border and in the Oslo fjord in the lower left corner.
The contour lines are similar for all three grid-resolutions, and
the particle velocities (right column) are in general consistent with the reference solution from NorKyst-800.
Due to the scale of the domain, it is harder to compare the fine details in Figure~\ref{fig:complete_coast_23_hours}, but the large scale features from NorKyst-800 are found in both the high and original grid resolutions.
As we have seen in the other cases, the low-resolution simulation maintains less details.
It should be noted that by zooming into the same areas simulated in Case 1 and 2, we have seen that the results from running the complete domain are similar within each resolution as to the results shown in Figures~\ref{fig:norwegian_sea_23_hours} and \ref{fig:lofoten_23_hours}.

\input{timing_results.tex}
Table~\ref{tab:timing_norway} reports the run time for simulating this case on a laptop, a desktop, and a server GPU.
The laptop contains a dedicated GeForce 840M GPU, which is powerful for a laptop but on the low end of the GPU performance spectrum.
It completes the simulation in slightly less than 54~minutes for the original grid resolution of NorKyst-800, which is comparable to the time required for 256 CPUs to run the NorKyst-800 simulation.
On a desktop with a slightly old GeForce 780GTX gaming GPU, the same simulation is carried out in approximately 10~minutes.
Tesla P100 is a GPU often found in supercomputers, thus belonging to the very high-end of the performance (and price) spectrum, and runs the 23-hour forecast in approximately 3~minutes.
Note that for the high-resolution scheme, the time step is on average just 0.46~seconds, meaning that the results shown in Figures~\ref{fig:norwegian_sea_23_hours}--\ref{fig:complete_coast_23_hours} are obtained after 180~000 time steps, using only single-precision floating-point operations.

\subsection{Tidal forecast verification and validation}
The oceanographic phenomena best captured and maintained by the shallow-water equations are arguably tides and storm surges.
We therefore use Case~3 to generate tidal forecasts for five locations along the Norwegian coast (see Figure~\ref{fig:case_locations}) and compare these with the hourly values from the NorKyst-800 dataset.
The locations have been chosen based on the availability of actual sea-level observation stations, and we use these to show the realism in the forecasts generated by both NorKyst-800 and the current model.
It should be noted that the sea level variation depends on more physical processes than those modeled by NorKyst-800 and our code, and we therefore expect a discrepancy between the forecasts and the observations. 
For us, it is therefore most relevant to use the NorKyst-800 result as a reference solution.
The five locations (see Figure~\ref{fig:case_locations}) are Bergen and Honningsv{\aa}g close to the open sea; Oslo and Narvik located in the innermost parts of long fjords; and Bod{\o}, which is affected by how the Lofoten archipelago catches the tidal waves.

\newcommand{\slpwidth}{0.85}
\begin{figure}[t!]
    \captionsetup[subfloat]{farskip=0pt,captionskip=0pt}
    \centering
    \subfloat{
        \includegraphics[width=\slpwidth\linewidth, trim=0 1.45cm 0 0.2cm, clip]{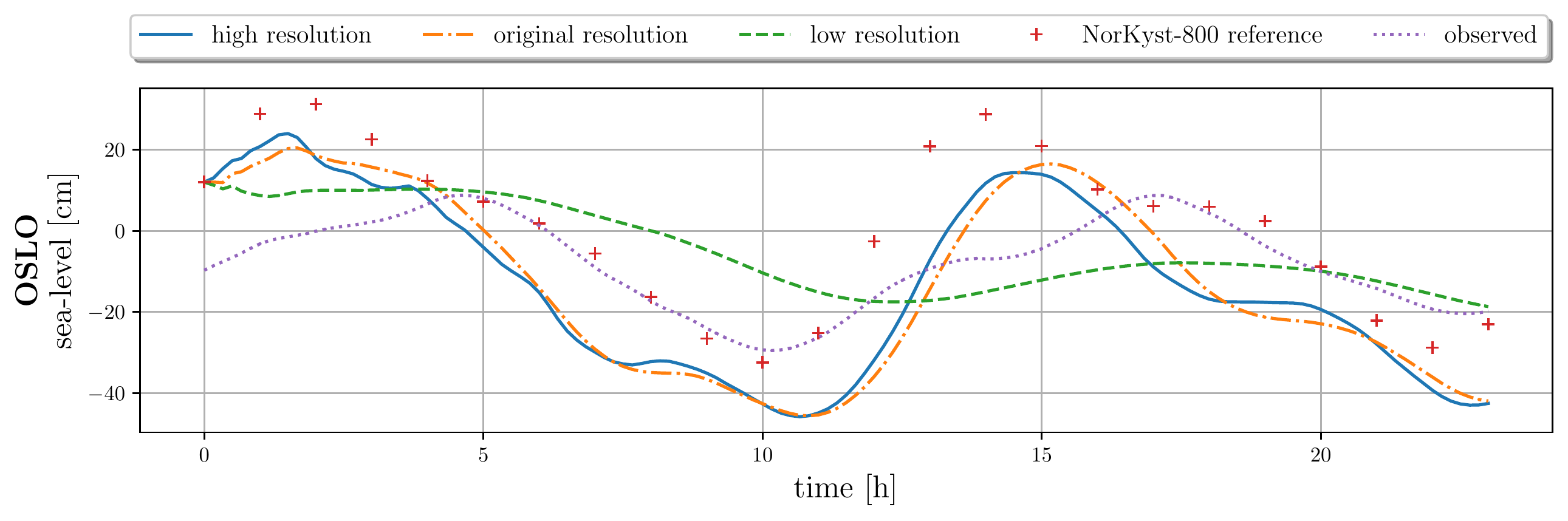}
    }
    \\
	\subfloat{
	    \includegraphics[width=\slpwidth\linewidth, trim=0 1.45cm 0 1.45cm, clip]{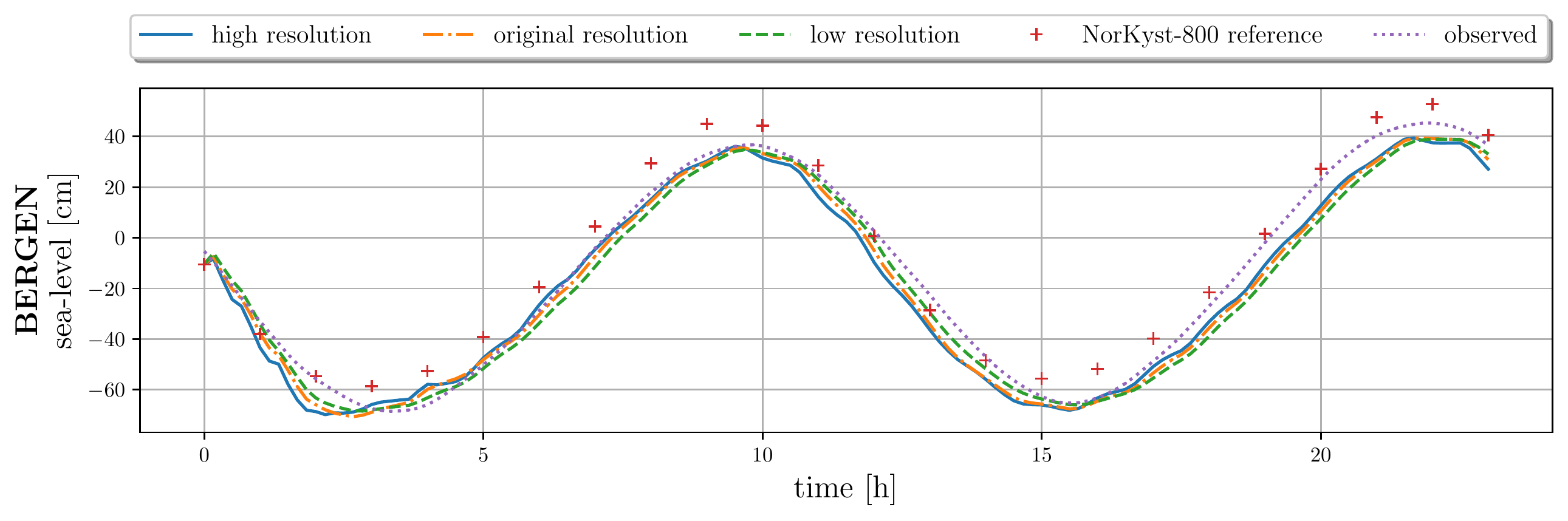}
	}
    \\
	\subfloat{
	    \includegraphics[width=\slpwidth\linewidth, trim=0 1.45cm 0 1.45cm, clip]{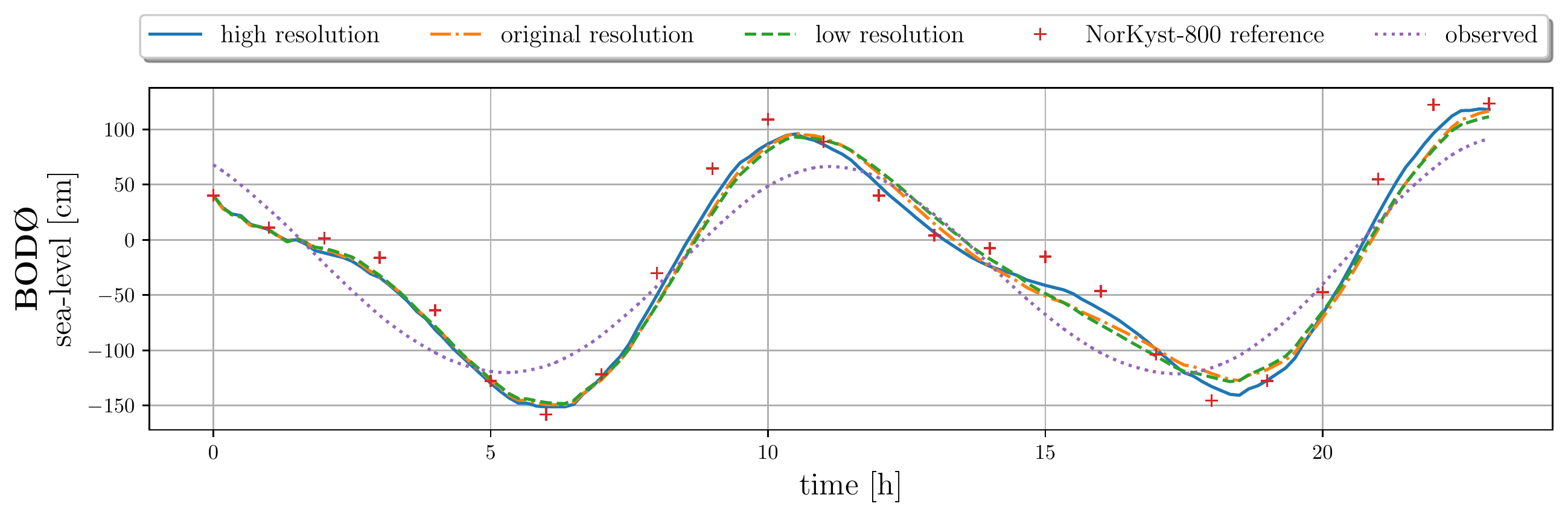}
	}
    \\
    \subfloat{
        \includegraphics[width=\slpwidth\linewidth, trim=0 1.45cm 0 1.45cm, clip]{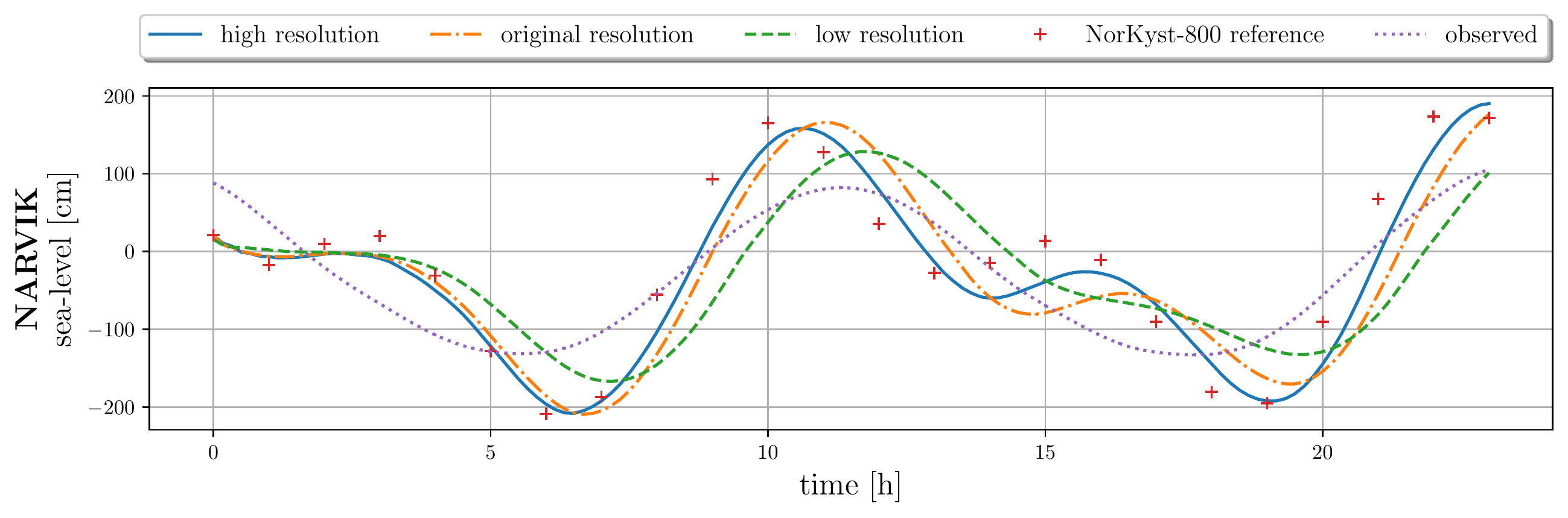}
    }
    \\
    \subfloat{
        \includegraphics[width=\slpwidth\linewidth, trim=0 0.2cm 0 1.45cm, clip]{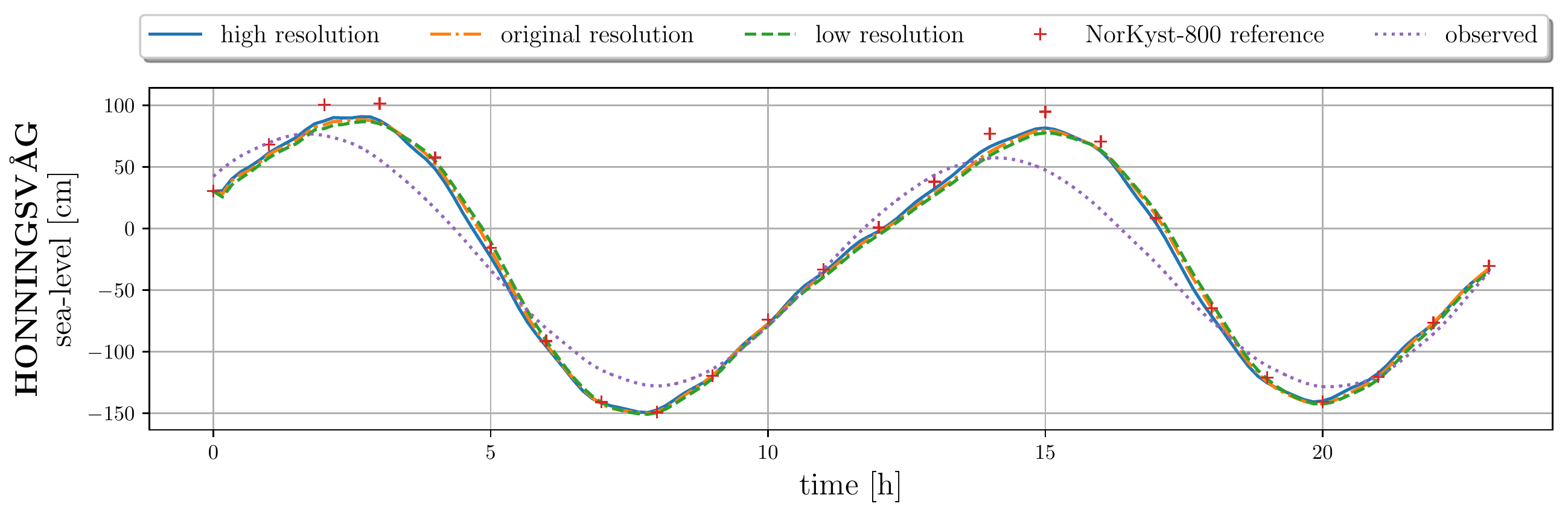}
    }
\caption{Tidal predictions, reference NorKyst-800 results, and official gauge measurements at selected locations generated by the simulations discussed in Case 3. 
    All grid resolutions capture the tide as predicted by NorKyst-800 with only minor discrepancies at coastal locations, such as Bergen, Bod{\o} and Honningsv{\aa}g. 
    The grid resolution comes much more into play deep inside the fjords, such as at Oslo and Narvik. 
    Here, the low-resolutions grid no longer manages to represent the bathymetry well enough for the tide to enter the fjord correctly (especially in the very narrow Oslo fjord), but we still get fair results with the high-resolution grid.
    The weakly dotted line is actual observed sea-surface level at the respective locations in the time range of the forecast.
    It should be noted that official tidal forecasts have additional physical terms, which is the main reason why the NorKyst-800 is off.
    }
    \label{fig:tidal_predictions}
\end{figure}

Figure~\ref{fig:tidal_predictions} shows the tidal forecasts generated from Case 3 for each of the five locations, sorted from south to north.
First of all, we see that our forecasts at Honningsv{\aa}g are very well in accordance with NorKyst-800.
The same is the case in Bergen, even though our simulations here consistently give slightly too low values for $\eta$.
At Bod{\o}, we see that NorKyst-800 gives a delay in the transition from high to low tide between 14 and 16 hours, and we can see weak indications of this in our high-resolution simulation as well.
All these locations are at the coast, which makes the tides here relatively easy to capture, but it should still be noted that they are all protected by islands.
The tides in Oslo and Narvik are seen to be harder to capture, as these two cities are located at the innermost parts of two long fjords, and we get larger differences between the different grid resolutions.
In Narvik, we get a large improvement by using the high-resolution grid over the original and low resolutions, but overall, our results are not too far from the reference solution.
Oslo is a particularly hard case, as it is positioned within a very narrow and shallow fjord.
This leads to larger relative discrepancies here, even though the absolute difference is roughly the same as in Narvik.
We see that the shape of our high-resolution result is a bit different than for the reference solution.
The tidal signal predicted by the low-resolution grid in Oslo is very weak, as the fjord almost becomes closed off at this resolution.
Finally, note that our solutions are well behaved already from the start, meaning that we manage to initialize our simulations in a balanced state.

\section{Summary}
\label{sec:summary}
We have presented a GPU simulation framework suitable for real-world oceanographic applications.
The framework is based on a modern high-resolution finite-volume method for the shallow-water equations~\cite{Chertock2017} with new adaptations and extensions to handle moving wet-dry fronts, land mask, bottom friction, wind forcing, varying north vector, varying latitude, and external boundary conditions.
The numerical algorithm is also improved to facilitate for efficient implementation on GPUs using single-precision floating-point operations and includes an efficient reformulation of a problematic recursive term.
The framework is designed to be initialized from operational ocean forecasts issued by the Norwegian Meteorological Institute from a three-dimensional ocean model.

We have presented second-order grid convergence for a synthetic but challenging benchmark containing complex topography and varying north vector.
We have also shown that the computational performance follows the expected weak scaling.

Finally, we have validated the simulation framework through three different real-world cases along the Norwegian coastline, initialized from ocean forecasts produced by the NorKyst-800 operational model with an $800~\mathrm{m} \times 800~\mathrm{m}$ horizontal resolution.
Our results demonstrate that the framework manages to maintain fine rotational structures, especially when increasing the grid resolution to $400~\mathrm{m} \times 400~\mathrm{m}$.
On a coarse grid (1600~m grid cells), however, the features are lost and the simulations get dominated by grid effects.
The tidal forecasts show that we are able to predict the observed tides relatively well at coastal locations on any of the three grid resolutions used, when compared to the NorKyst-800 reference data.
We see, however, that within long fjords, our results are improved when increasing the grid resolution, compared to using the original horizontal resolution of the NorKyst-800 model.

\subsection*{Acknowledgements}
This research has been funded by the Research Council of Norway under grant number 250935 (GPU Ocean). The GPU Ocean core project team consists of Göran Broström, Kai Christensen, Knut-Andreas Lie, and Martin Lilleeng Sætra, and the authors are sincerely grateful for their collaboration and discussions, which have influenced and enabled this work. The authors also thank {\O}yvind S{\ae}tra and Jon Albretsen for feedback on drafts of the manuscript. The GPU Ocean project has received support in form of compute time on UNINETT Sigma2 - the National Infrastructure for High Performance Computing and Data Storage in Norway under project number nn9550k. The authors declare that we have no competing interests related to this work.

\subsection*{CRedIT author statement}
A. R. Brodtkorb: Conceptualization, Methodology, Software, Investigation, Writing - Original Draft, Writing - Review \& Editing, Visualization, Project administration, Supervision, H. H. Holm: Conceptualization, Methodology, Software, Investigation, Writing - Original Draft, Writing - Review \& Editing, Visualization.

\bibliography{references}

\end{document}

%% file: tikz/tikzsetup.tex
\usetikzlibrary{calc,trees,positioning,arrows,chains,shapes.geometric,%
    decorations.pathreplacing,decorations.pathmorphing,shapes,%
    matrix,shapes.symbols,scopes,fit}

\tikzset{
  >=stealth,
  auto,
  x=4em,
  y=4em,
}
\tikzstyle{text_box}=[text centered, text width=8em, minimum height=2em]
\tikzstyle{round_rectangle}=[
  rectangle,
  rounded corners,
  thick,
]
\tikzstyle{yellow_tint}=[draw=yellow!50, fill=yellow!20]
\tikzstyle{blue_tint}=[draw=blue!50, fill=blue!20]
\tikzstyle{green_tint}=[draw=green!50, fill=green!20]
\tikzstyle{red_tint}=[draw=red!50, fill=red!20]
\tikzstyle{violet_tint}=[draw=violet!50, fill=violet!20]
\tikzstyle{orange_tint}=[draw=orange!50, fill=orange!20]
\tikzstyle{gray_tint}=[draw=black!30, fill=black!10]
\tikzstyle{cache}=[round_rectangle, text_box, orange_tint]
\tikzstyle{myarrow}=[->, thick]


%% file: tikz/equation.tikz
\begin{tikzpicture}[scale=0.5]
  \path[fill=blue!30]
    (0, 0) --
    (0,1.7) .. controls +(1,-1) and +(-2,1) .. (5,1.6) -- 
    (5, 0) -- (0, 0);
  \draw (0,1.7) .. controls +(1,-1) and +(-2,1) .. (5,1.6);


  \foreach \y in {0.2, 0.5, 0.8, 1.1, 1.4, 1.7}
    \draw[thick,->] (4.3, \y) -- +(-0.5, 0);
  \node[anchor=west,inner sep=1pt] at (4.3, 1.25) {$hu$};

  \draw[thick,<->] (3.4, 0.0) -- (3.4, 1.89) node[midway,inner sep=1pt] {$h$};

  \draw[thin] (0, 0) -- +(5, 0)
    (0, -0.1) -- (0, 0.1)
    (5, -0.1) -- (5, 0.1);
\end{tikzpicture}

%% file: tikz/equation_discrete.tikz
\begin{tikzpicture}[scale=0.5]
  \path[fill=blue!30] (0, 0) --
    (0, 0) -- 
    (0, 1.45) -- (1, 1.45) -- 
    (1, 1.5) -- (2, 1.5) -- 
    (2, 1.7) -- (3.0, 1.7) -- 
    (3, 1.9) -- (4.0, 1.9) -- 
    (4, 1.8) -- (5.0, 1.8) -- 
    (5.0, 0) -- (0, 0)
    ;


  \draw[thick] (0,1.45) -- (1, 1.45) (0.5, 1.45) circle (0.05)
    (1, 1.5) -- (2, 1.5) (1.5, 1.5) circle (0.05)
    (2, 1.7) -- (3.0, 1.7) (2.5, 1.7) circle (0.05)
    (3, 1.9) -- (4.0, 1.9) (3.5, 1.9) circle (0.05)
    (4, 1.8) -- (5.0, 1.8) (4.5, 1.8) circle (0.05)
    ;


  \node[right] (w) at (1.2, 2.1) {$Q_i$};
  \draw[thick,->] (w.east) .. controls +(0.2, 0) and +(-0.1, 0.2) .. (2.47, 1.75);

  \draw[thin] (0, 0) -- +(5, 0)
    (0, -0.1) -- (0, 0.1)
    (1, -0.1) -- (1, 0.1)
    (2, -0.1) -- (2, 0.1)
    (3, -0.1) -- (3, 0.1)
    (4, -0.1) -- (4, 0.1)
    (5, -0.1) -- (5, 0.1);

\end{tikzpicture}

%% file: tikz/equation_slopes.tikz
\begin{tikzpicture}[scale=0.5]
  \path[fill=blue!30] 
    (0, 0) -- 
    (0, 1.45) -- (1, 1.45) -- 
    (1, 1.45) -- (2, 1.55) -- 
    (2, 1.6) -- (3.0, 1.8) -- 
    (3, 1.9) -- (4.0, 1.9) -- 
    (4, 1.9) -- (5.0, 1.7) -- 
    (5.0, 0) -- (0, 0)
    ;


  \draw[thick] (0,1.45) -- (1, 1.45) (0.5, 1.45) circle (0.05)
    (1, 1.45) -- (2, 1.55) (1.5, 1.5) circle (0.05)
    (2, 1.6) -- (3.0, 1.8) (2.5, 1.7) circle (0.05)
    (3, 1.9) -- (4.0, 1.9) (3.5, 1.9) circle (0.05)
    (4, 1.9) -- (5.0, 1.7) (4.5, 1.8) circle (0.05)
    ;

  \node[right] (w) at (0.2, 2.1) {$\partial Q / \partial x$};
  \draw[thick,->] (w.east) .. controls +(0.2, 0) and +(-0.1, 0.2) .. (2.3, 1.73);

  \draw[thin] (0, 0) -- +(5, 0)
    (0, -0.1) -- (0, 0.1)
    (1, -0.1) -- (1, 0.1)
    (2, -0.1) -- (2, 0.1)
    (3, -0.1) -- (3, 0.1)
    (4, -0.1) -- (4, 0.1)
    (5, -0.1) -- (5, 0.1);

\end{tikzpicture}

%% file: tikz/equation_point_values.tikz
\begin{tikzpicture}[scale=0.5]
  \path[fill=blue!30]
    (0, 0) -- 
    (0, 1.7) -- (1, 1.2) -- 
    (1, 1.45) -- (2, 1.55) -- 
    (2, 1.6) -- (3.0, 1.8) -- 
    (3, 1.9) -- (4.0, 1.9) -- 
    (4, 1.9) -- (5.0, 1.7) -- 
    (5.0, 0) -- (0, 0)
    ;
  \draw[thick] (0.05,1.7) circle (0.05) (0.95, 1.2) circle (0.05)
    (1.05, 1.45) circle (0.05) (1.95, 1.55) circle (0.05)
    (2.05, 1.6) circle (0.05) (2.95, 1.8) circle (0.05)
    (3.05, 1.9) circle (0.05) (3.95, 1.9) circle (0.05)
    (4.05, 1.9) circle (0.05) (4.95, 1.7) circle (0.05)
    ;


  \node[right,inner sep=1pt] (w) at (1.5, 2.1) {$Q^-$};
  \draw[thick,->] (w.east) .. controls +(0.2, 0) and +(-0.2, 0.2) .. (2.85, 1.85);

  \node[right,inner sep=1pt] (w) at (3.7, 2.3) {$Q^+$};
  \draw[thick,->] (w.west) .. controls +(-0.2, 0) and +(0.2, 0.2) .. (3.15, 1.95);

  \draw[thin] (0, 0) -- +(5, 0)
    (0, -0.1) -- (0, 0.1)
    (1, -0.1) -- (1, 0.1)
    (2, -0.1) -- (2, 0.1)
    (3, -0.1) -- (3, 0.1)
    (4, -0.1) -- (4, 0.1)
    (5, -0.1) -- (5, 0.1);
\end{tikzpicture}

%% file: benchmark_results.tex
\begin{table*}
\caption{Performance for our simulator. The simulation time is approximately eight times larger when we double the resolution. This is because we have four times as many cells and twice as many timesteps to reach the same end time. The download increases by a factor four, which is to be expected as we transfer four times as many data elements.}
\label{tab:performance}
\small
\begin{center}
\begin{tabularx}{0.6\linewidth}{crrrr}
	\textbf{Resolution} & \textbf{Simulation time} & \textbf{Factor} & \textbf{Download time} & \textbf{Factor} \\ 
$512^2$ & 0.169710 & - & 0.000929& -\\
$1024^2$ & 1.170814 &6.9 & 0.002773 &3.0\\
$2048^2$ & 9.133564 &7.8 & 0.019642 &7.1\\
$4096^2$ & 72.832028 &8.0 & 0.077541 &3.9\\
$8192^2$ & 582.599290 &8.0 & 0.305405 &3.9\\
$16384^2$ & 4695.904546 &8.0 & 1.218880 &4.0\\
\end{tabularx}
\end{center}
\end{table*}

%% file: timing_results.tex
\renewcommand{\tabularxcolumn}[1]{b{#1}}

\begin{table*}[b]
\caption{Timing results different resolutions of running the complete Norwegian coast (Case 3)  with the original, high and low resolution on different computer types.}
\label{tab:timing_norway}
\small 
\begin{center}
\newcolumntype{L}{>{\raggedright\arraybackslash}X}%
\newcolumntype{R}{>{\raggedleft\arraybackslash}X}%
\begin{tabularx}{0.8\linewidth}{lrrRRR}
	\textbf{Resolution} & \textbf{Num cells} & \textbf{Average $\Delta t$} &
            	\textbf{Laptop GeForce~840M Maxwell~(2014)}  &  
            	\textbf{Desktop GeForce~GTX780 Kepler~(2013) } & 
            	\textbf{ $\quad$ Server Tesla~P100 Pascal~(2016)}  \\ 
    \textbf{Low}        &   541 875 & 1.85~s &     9m 22s &     1m 29s & 47s \\    
    \textbf{Original}   & 2 167 500 & 0.92~s &    53m 49s &    10m 42s & 3m 13s \\
    \textbf{High}       & 8 670 000 & 0.46~s & 6h 10m 09s & 1h 23m 22s & 23m 21s \\      
\end{tabularx}
\end{center}
\end{table*}

